\documentclass{aa} 
\usepackage{times,amssymb,amsmath, epsfig,graphicx} 

\newcommand{\chandra}{{\it Chandra} }
\newcommand{\xmm}{{\it XMM-Newton} }

\newcommand{\sax}{{\it BeppoSAX} }
\newcommand{\NH}{$\rm{N_{H}}\,$}

\title{\xmm observation of the interacting cluster Abell 3528}

\author{F. Gastaldello \inst{1,2}
  \and S. Ettori \inst{3}
  \and S. Molendi \inst{1}
  \and S. Bardelli \inst{4}
  \and T. Venturi  \inst{5}
  \and E. Zucca \inst{4}
  }

\institute{
  IASF - CNR Sezione di Milano ``G.Occhialini'', 
  Via Bassini 15, I-20133  Milano, Italy 
  \and
  Universit\`a di Milano Bicocca, Dip. di fisica, 
  P.za della Scienza 3, I-20126 Milano, Italy 
  \and
  European Southern Observatory,  
  Karl-Schwarzschild-Stra\ss e 2,
  85748, Garching bei M\"unchen, Germany
  \and
  INAF - Osservatorio Astronomico di Bologna, 
  via Ranzani 1, I-40127 Bologna, Italy
  \and
  Istituto di Radioastronomia del CNR,
  via Gobetti 101, I-40129, Bologna, Italy 
  }
\offprints{}
\mail{gasta@mi.iasf.cnr.it}

\date{Received  / Accepted }

\begin{document}
\titlerunning{\xmm analysis of A3528}

%%%%%%%%%%%%%%%%%%%%%%%%%%%%%%%%%%%%%%%%%%%%%%%%%%%%%%%%%%%%%%%%%%%%%%%%%%%%%%

\abstract{We analyze the \xmm dataset of the interacting cluster of galaxies
Abell 3528 located westward in the core of the Shapley Supercluster, the
largest concentration of mass in the nearby Universe.
A3528 is formed by two interacting clumps (A3528-N at North 
and A3528-S at South) separated by 0.9 $h_{70}^{-1}$ Mpc
at redshift 0.053. \xmm data describe these clumps as relaxed structure with
an overall temperature of $4.14 \pm 0.09$ and $4.29 \pm 0.07$ keV in
A3528-N and A3528-S, respectively, and a core cooler by a factor 
1.4--1.5 and super-solar metal abundance in the inner 30 arcsec.
These clumps are connected by a X-ray soft, bridge-like emission
and present asymmetric surface brightness with significant excess 
in the North--West region of A3528-N and in the North--East area
of A3528-S. 
However, we do not observe any evidence of shock heated gas, both
in the surface brightness and in the temperature map.
Considering also that the optical light distribution is more 
concentrated around A3528-N and makes A3528-S barely detectable,  
we do not find support to the originally suggested head-on 
pre-merging scenario and conclude that A3528 is in a off-axis 
post-merging phase, where the closest cores encounter happened 
about 1--2 Gyrs ago.
}
%%%%%%%%%%%%%%%%%%%%%%%%%%%%%%%%%%%%%%%%%%%%%%%%%%%%%%%%%%%%%%%%%%%%%%%%%%%%%%

\maketitle

%%%%%%%%%%%%%%%%%%%%%%
\section{Introduction}
%%%%%%%%%%%%%%%%%%%%%%

Clusters of galaxies are thought to form by accretion and merging of
subunits in a hierarchical bottom-up scenario. Numerical simulations on
scales of cosmological relevance revealed that mergings happen along
preferential directions, called density caustics, which define matter flow 
regions, at whose intersection rich clusters are formed (Cen \& Ostriker 1994,
 Colberg et al. 1999).    
Cluster mergings are therefore the most common and energetic phenomena in
the Universe, leading to an energy release of as much as $10^{64}$ ergs on
a timescale of the order of Gyrs, whose astrophysical consequences are expected
to be both thermal (presence of shocks, changes in the temperature and gas
distribution of the Intra Cluster Medium) and non thermal 
(appearance of radio sources like Halos or Relics) as well as influences
on the galaxy population (presence of starbust galaxies, wide angle tail
radiosources) (see Sarazin 2002, Girardi \& Biviano 2002, Buote 2002, B\"ohringer \& Schuecker 2002, Forman et al. 2002, Feretti \& Venturi 2002, Giovannini \& Feretti 2002).
The understanding of cluster merging thus needs a full multi wavelength 
analysis (optical, radio and X-ray) in order to have a more 
global view of the phenomenon and the 
improvement and extension of available data.
For the second issue the advent of the new generation of X-ray observatories 
is giving a lot of results, as \chandra reveled with the long
seeked first clear example of bow shock in the ICM (Markevitch et al. 2002), 
the most prominent feature seen in cluster merging simulations 
(see Schindler 2002 for a review), and the 
discovery of the completely unexpected phenomenon of cold fronts, which are
direct consequences of the survival of cluster cores during the process of
merging, at least in the most spectacular cases of A3667 and A2142 
(Vikhlinin et al. 2001, Forman et al. 2002 for a review).      

In the same way as the caustics seen in the simulations, rich superclusters
are the ideal environment for the detection of cluster mergings, because the
peculiar velocities induced by the enhanced local overdensity (of 
the order of $\sim 10$) of the large-scale structure favor the 
cluster-cluster collisions.    
The Shapley Supercluster (Shapley 1930) is the largest concentration 
of mass within $z=0.1$ (Zucca et al. 1993, Bardelli et al. 1994, 
Ettori, Fabian \& White 1997) with a high local overdensity (of the
order of $\sim 10$ on scales of $10\,\rm{h}^{-1}$ Mpc, Bardelli et al. 2000)
showing remarkable examples of cluster mergings at various evolutionary stages.
In particular, two cluster complexes are found  in this
concentration, around A3558 in the core and A3528 westward,
respectively.
These structures, whose spatial scales are of order of
$\sim\,5\,\rm{h}^{-1}$ Mpc (Bardelli et al. 2000) are formed by strongly 
interacting clusters. The complex dominated by A3558 is probably a merging
seen just after the first core-core encounter (Bardelli et al. 1998, Hanami et al. 1999). 
The other complex is formed 
by the ACO (Abell, Corwin \& Olowin 1989) clusters A3528, A3530 and A3532, 
located northwest of the A3558 complex.
\newline   
In this paper we concentrate on the \xmm observation of the cluster A3528.

The outline of the paper is as follows. We describe in Sect.~2 
the general properties of A3528 and present in Sect.~3 the X-ray spatial
and spectral analysis of the \xmm observation. In Sect.~4, we consider
the state of the merging how it appears at optical and radio 
wavelengths. Finally, we discuss and summarize our results 
in Sect.~5 and Sect.~6, respectively. More details on technical aspects of 
the X-ray analysis are presented in the Appendices at the end of the paper.

At the nominal redshift of A3528 (z=0.053), 1 arcmin corresponds to 
62 kpc ($H_0 =$ 70 $h_{70}$ km s$^{-1}$ Mpc$^{-1}$,
$\Omega_{\rm m} = 1 - \Omega_{\Lambda} =$ 0.3).
In the following analysis, all the quoted errors are
at $1 \sigma$ (68.3 per cent level of confidence)
unless stated otherwise.

%%%%%%%%%%%%%%%%%%%%%%%%%%%
\section{The cluster A3528}
%%%%%%%%%%%%%%%%%%%%%%%%%%%

The cluster A3528 is a Bautz-Morgan type II galaxy cluster with 
richness of class 1, the highest in the complex being the other two
clusters, A3530 and A3532, of richness class 0.
Raychaudhury et al. (1991) found, using Einstein IPC, that this cluster is 
actually double, formed by two subcomponents, A3528-N and A3528-S, centered on 
the two dominant galaxies clearly visible in the optical, with a 
likely although poor indication of the presence of a cool core.
From a ROSAT pointed PSPC observation, Schindler (1996) showed evidence
of interaction, dividing the two subclusters in four semicircles, with the
semicircles facing the other subcluster with higher temperatures than the outer
ones, suggesting the presence of a shock. 
Henriksen \& Jones (1996) estimated a temperature of $2.7\pm0.8$ and
$2.9\pm0.9$ keV for A3528-N and A3528-S, respectively, suggesting as
unlikely the presence of a cool core in the two subclusters.
However this observation is not optimal: the cluster is off-centered respect
to the ROSAT pointing and partly obscured by the supporting ribs 
(see Fig.8 of Reid et al. 1998) and 
there are some concerns about the capability of the PSPC
to obtain accurate temperature determination (Markevitch \& Vikhlinin 1997, 
Ettori et al. 2000).

Optical (Bardelli et al. 2001, Baldi et al. 2001) and radio observations 
(Reid et al. 1997, Venturi et al. 2001) seem to indicate that the A3528 complex
is in a pre-merging phase because the cluster galaxies in the inner regions
are not far from the virial equilibrium and their optical and radio properties
are not yet affected from merging effects.

Donnelly et al. (2001) using ASCA GIS data found an overall cluster temperature
(comprising the two subclusters) of $4.7\pm0.3$ keV and analyzing five 
indicative regions (two regions for each subcluster, one around the core and 
one in the outer region, and a region between the two subclusters, see their 
Fig.1) provided further support to the idea that A3528 is in the early 
stages of a merger, where
the gas in both subclusters is only just beginning to interact. They claimed
(still by ROSAT data) that the cores show evidence of cool gas, the intensity 
contours are still azimuthally symmetric and the temperature of the gas located
between the subclusters is only marginally hotter ($\sim 15\%$) than 
the overall average for the cluster.   

%%%%%%%%%%%%%%%%%%%%%%%%
\section{X-ray analysis}
%%%%%%%%%%%%%%%%%%%%%%%%

%%%%%%%%%%%%%%%%%%%%%%%%%%%%%%%%%%%%%%%%%%%%%
\subsection{Observation and data preparation}
%%%%%%%%%%%%%%%%%%%%%%%%%%%%%%%%%%%%%%%%%%%%%

\begin{figure}[!]
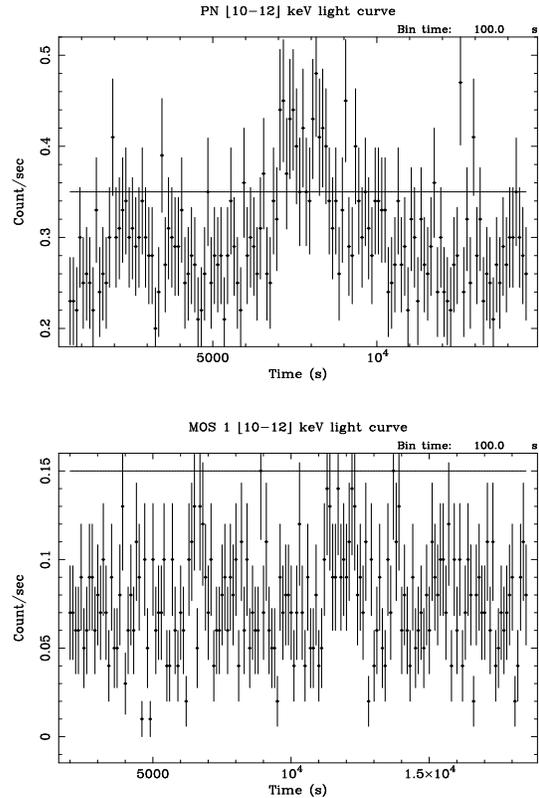

\begin{center} 
\epsfig{file=pn_lc.ps,width=5.0cm,height=7.0cm,angle=-90}
\vskip 0.5truecm
\epsfig{file=m1_lc.ps,width=5.0cm,height=7.0cm,angle=-90}
\caption{{Light curves for PN (upper panel) and, for comparison, of the MOS 1 
(lower panel) in the band [10-12] keV and the intensity filters we use.}\label{spflare}} 
\end{center}
\end{figure}

A3528 was observed by \xmm during rev. 374 with the MOS and PN 
detectors in Full Frame Mode with medium filter, for an  
exposure time of 16.4 ks for MOS
and 14.0 ks for PN. We have obtained calibrated event files for the three EPIC
cameras with SASv5.3.3 using the tasks \emph{emproc} and \emph{epproc}. 
\newline
The removal of bright pixels and hot columns was done, especially for the PN,
in a conservative way applying the expression (FLAG ==0) to the extraction
of spectra and images.
\newline 
To reject the soft proton flares, we accumulate 
the light curve in the [10-12] keV band (see comparison in Fig.{\ref{spflare}}), where
the emission is dominated by particle induced background: 
while the MOS camera are not affected by 
flares (count rates are lower than our intensity threshold, fixed at 15 cts/100s),
we observe a flare in the PN detector with a count rate exceeding our limit 
of 35 cts/100s. Thus, we reject all the PN events related to this flare
deriving a total effective exposure time of 11.0 ksec.
The pattern selection was [0;12] for MOS cameras and singles events for the PN.  

\begin{figure}[!]
\begin{center} 
\epsfig{file=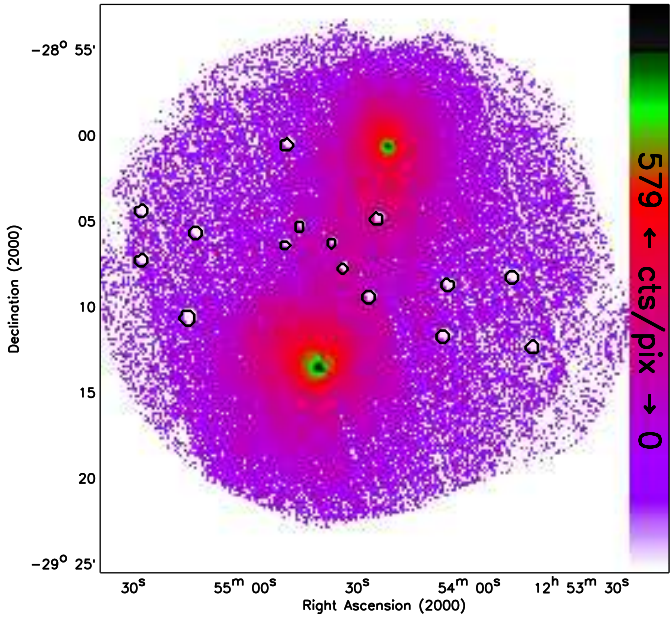,width=9.0cm,height=9.0cm}
\caption{{\xmm EPIC image of the cluster A3528 in the band [0.5-10.0] keV. 
The bright sources detected by the pipeline processing of the image are masked.
It is worth noticing the high concentration of these X-ray emitting
point sources in the region lying between the two clumps. None of these
have an associated redshift.
}\label{sources}} 
\end{center}
\end{figure}  

\begin{figure}[!]
\begin{center}
\epsfig{file=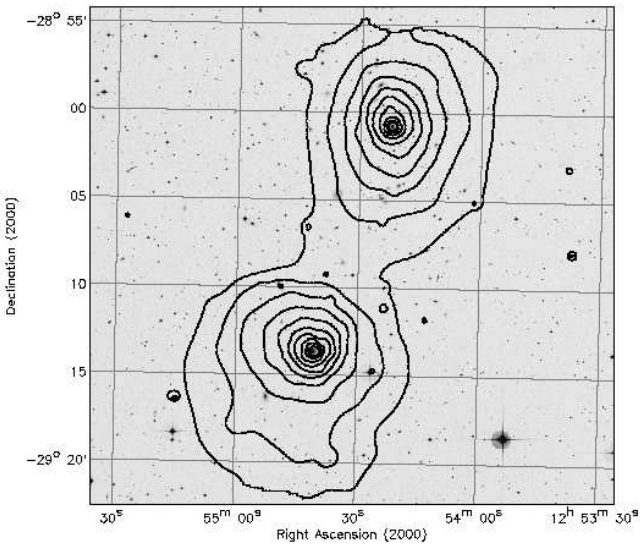,angle=-90,width=0.5\textwidth}
\caption{{\xmm contour plots of the EPIC image of the cluster A3528 
in the band [0.5-10.0] keV superposed on the DSS image, in the Bj band, 
of the same 
field of view. The contours are spaced by a factor of $\sim 1.4$ 
between 1.7 and 579 cts/pixel.
}\label{optical}}
\end{center}
\end{figure}

\begin{figure}[!]
\begin{center}
\epsfig{file=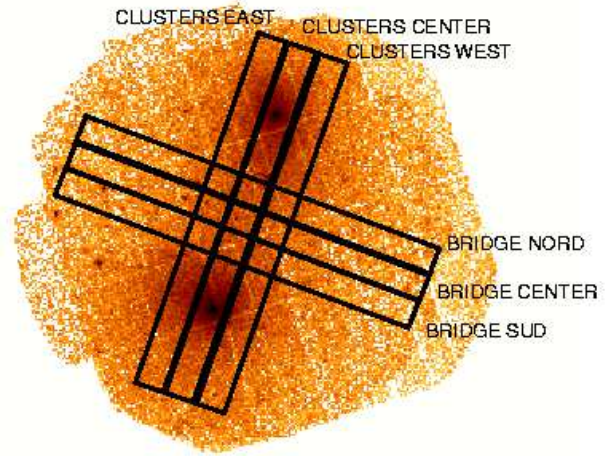,width=9.0cm,height=7.0cm}
\caption{{The regions used for the detection of the diffuse emission
between the two clumps are here superimposed on the total EPIC raw image.
The width of the strip is of 13.2 pixels $\approx 1.90$ arcmin.
The surface brightness profiles shown in Fig.~\ref{csmooth} are 
extracted along these strips and are obtained by
accumulating the counts in bins of 10 pixels ($\approx 1.44$ arcmin) each. 
The progression of the bins starts from north to south along
the ``clusters'' regions, while from west to east in the ``bridge'' regions.
}\label{clubridge}}
\end{center}
\end{figure}

To remove all other background components we do not use blank sky 
regions as the Lockman Hole observations or the background templates  
provided by the EPIC team (Lumb 2002), because all these fields are observed with
thin filter and with low levels of galactic absorption. We prefer to use
as background the long observation (100 ks) of the quasar APM 08279+5255 
(Hasinger et al. 2002) performed with medium filter, which has also a Galactic $\rm{N_{H}}$ of 
$4\times 10^{20}\,\rm{cm^{-2}}$ more similar to the one in the direction of A3528 
($6.1\times 10^{20}\,\rm{cm^{-2}}$). 
We performed the same selection criteria
for the background as the source, we removed all the bright sources and after rejection of 
soft proton flares we got an effective exposure time of 73 ks for MOS and 62 ks for PN.
We reprojected the background
field to the same sky coordinates of the source by means of the SAS task 
\emph{attcalc} and then performed the background subtraction in sky coordinates. 
  
\begin{figure*}[!]
\begin{center}
\hspace*{0.5cm}\hbox{ \epsfig{file=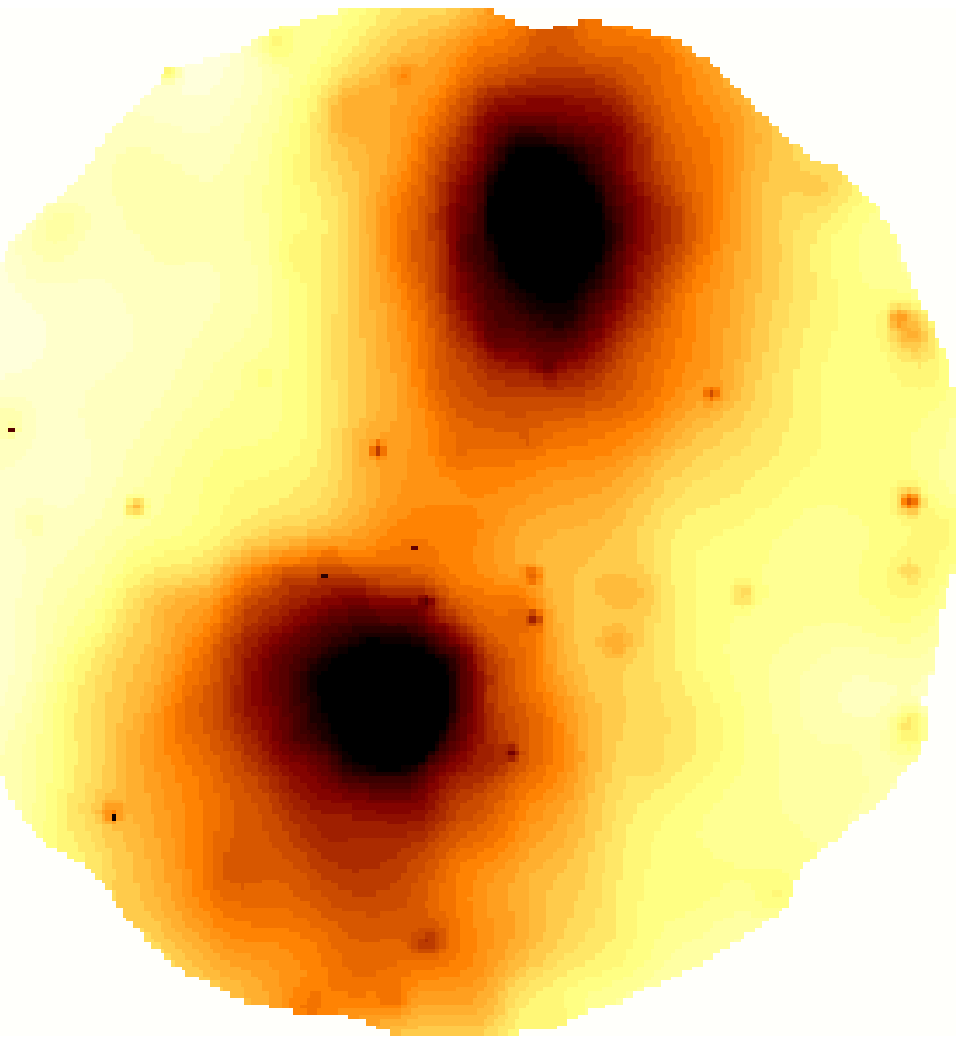,width=0.35\textwidth}
  \epsfig{file=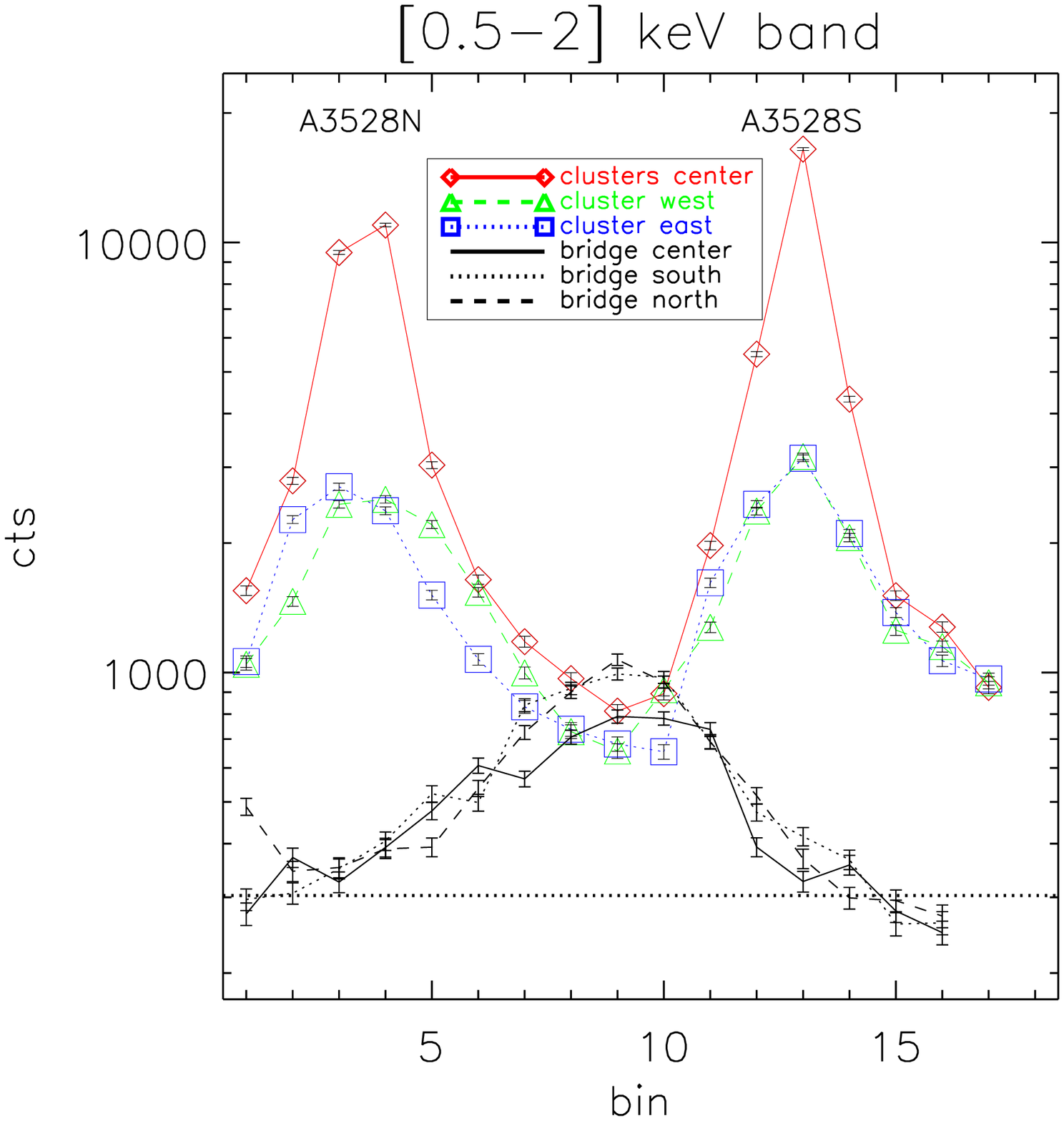,width=0.43\textwidth}
} \hspace*{0.5cm}\hbox{ \epsfig{file=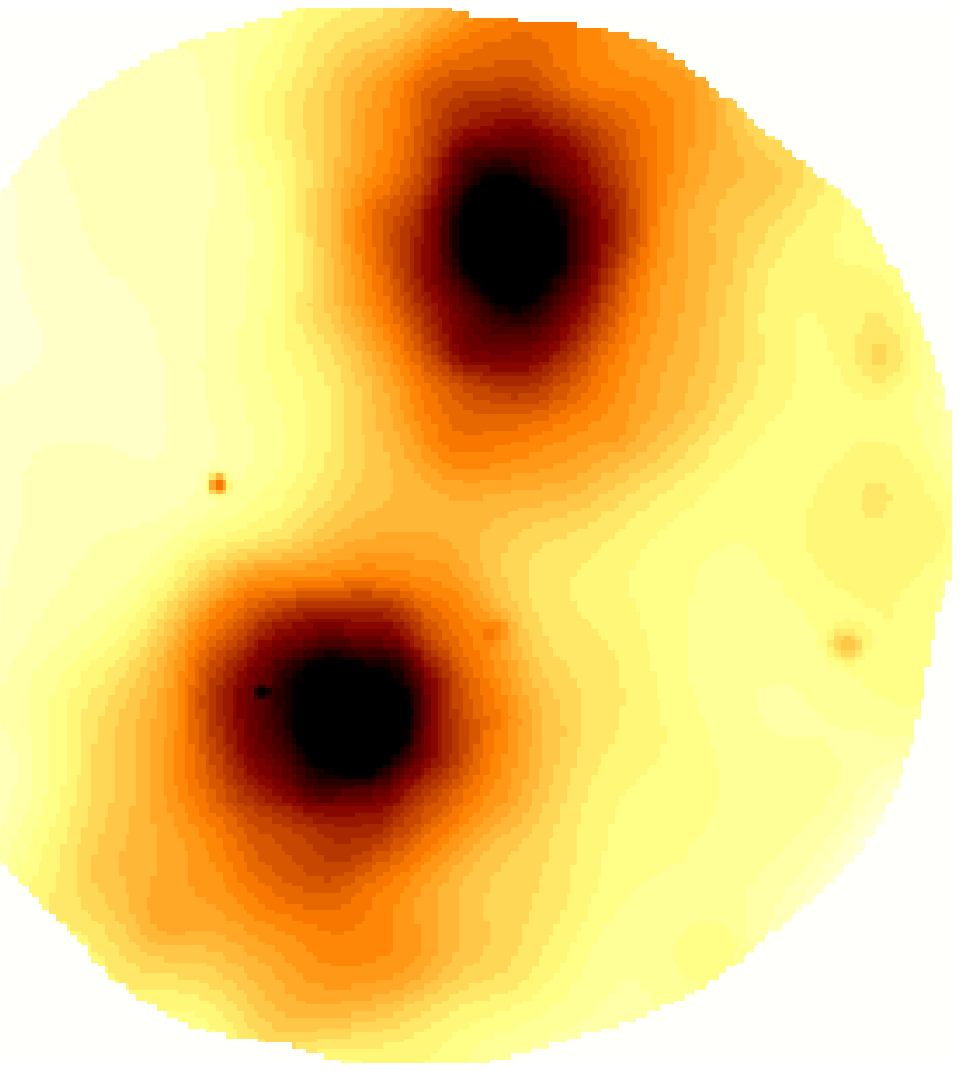,width=0.35\textwidth}
  \epsfig{file=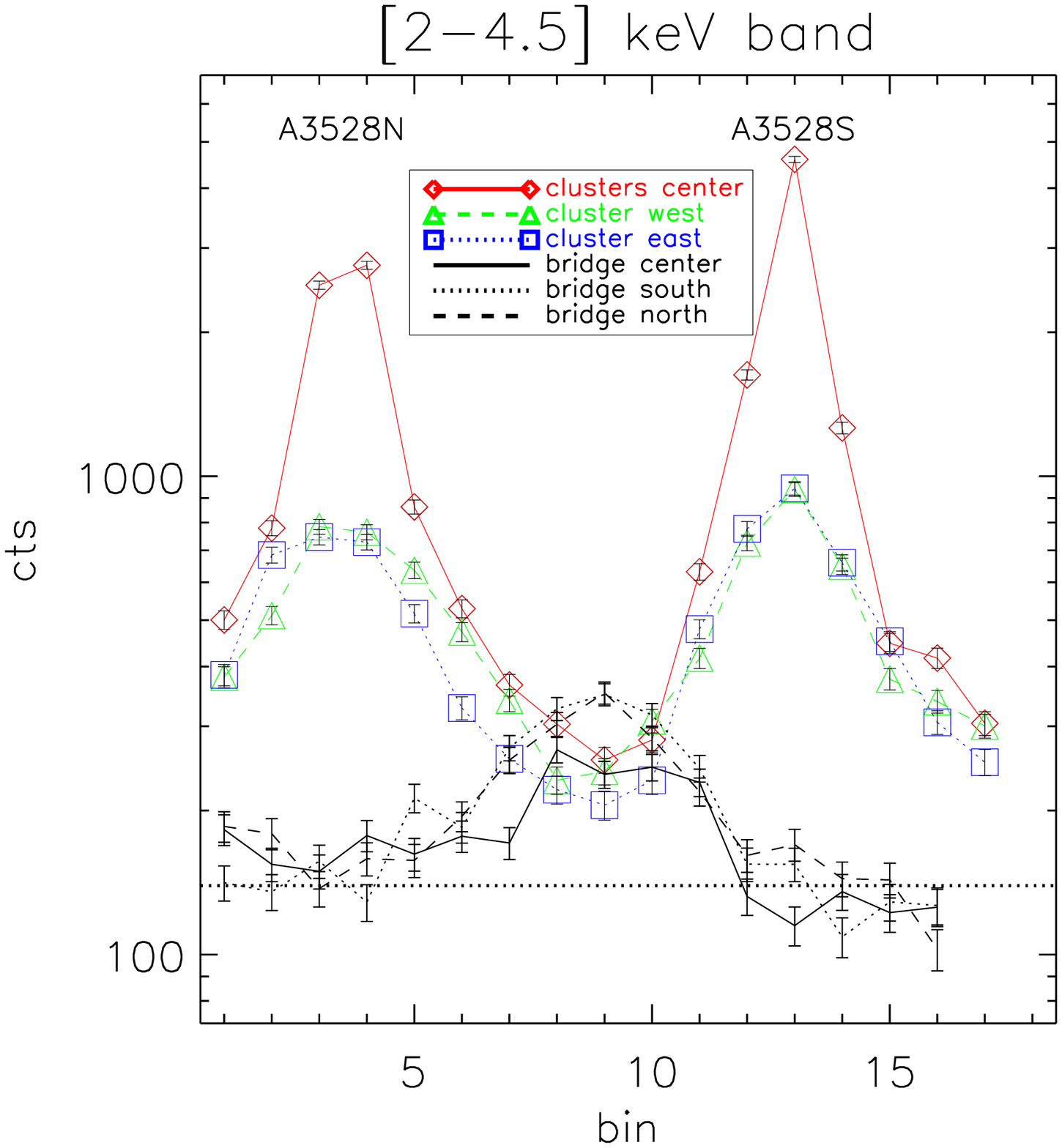,width=0.43\textwidth}
} \hspace*{0.5cm}\hbox{ \epsfig{file=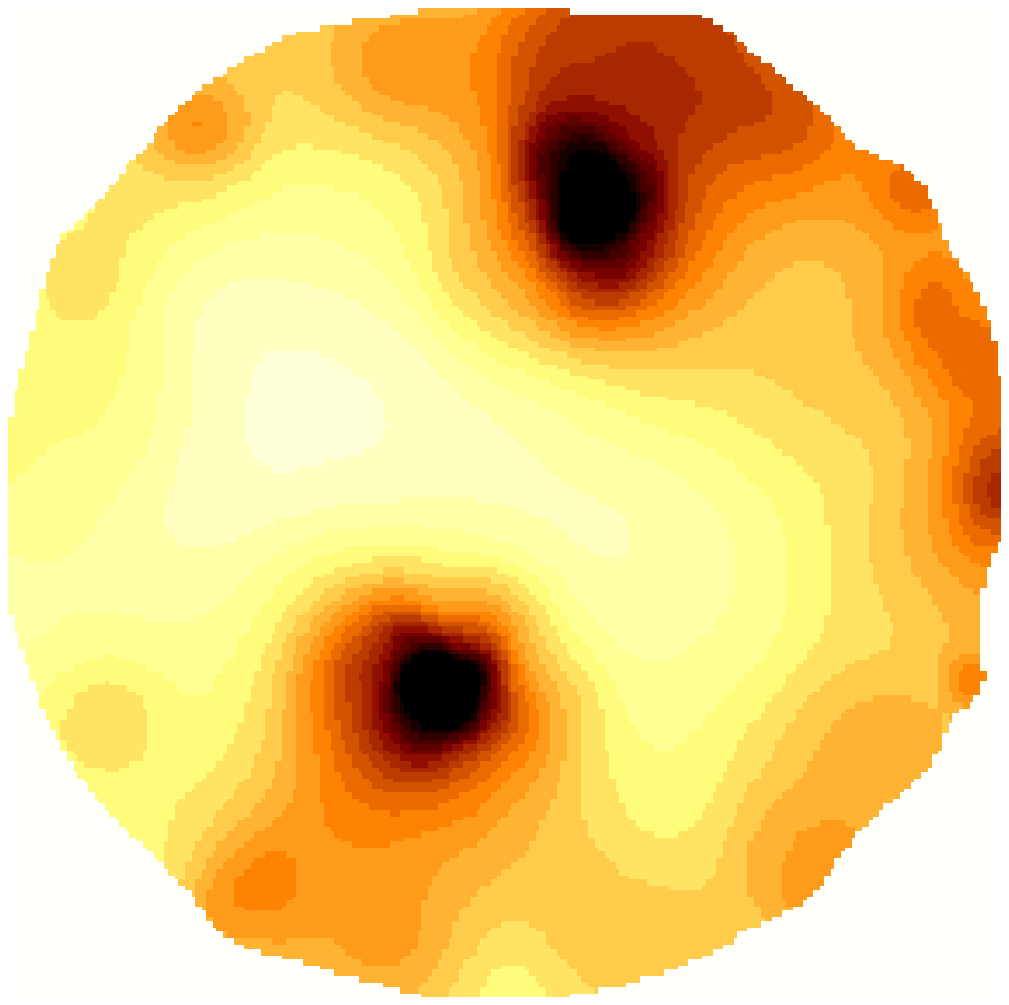,width=0.35\textwidth}
\epsfig{file=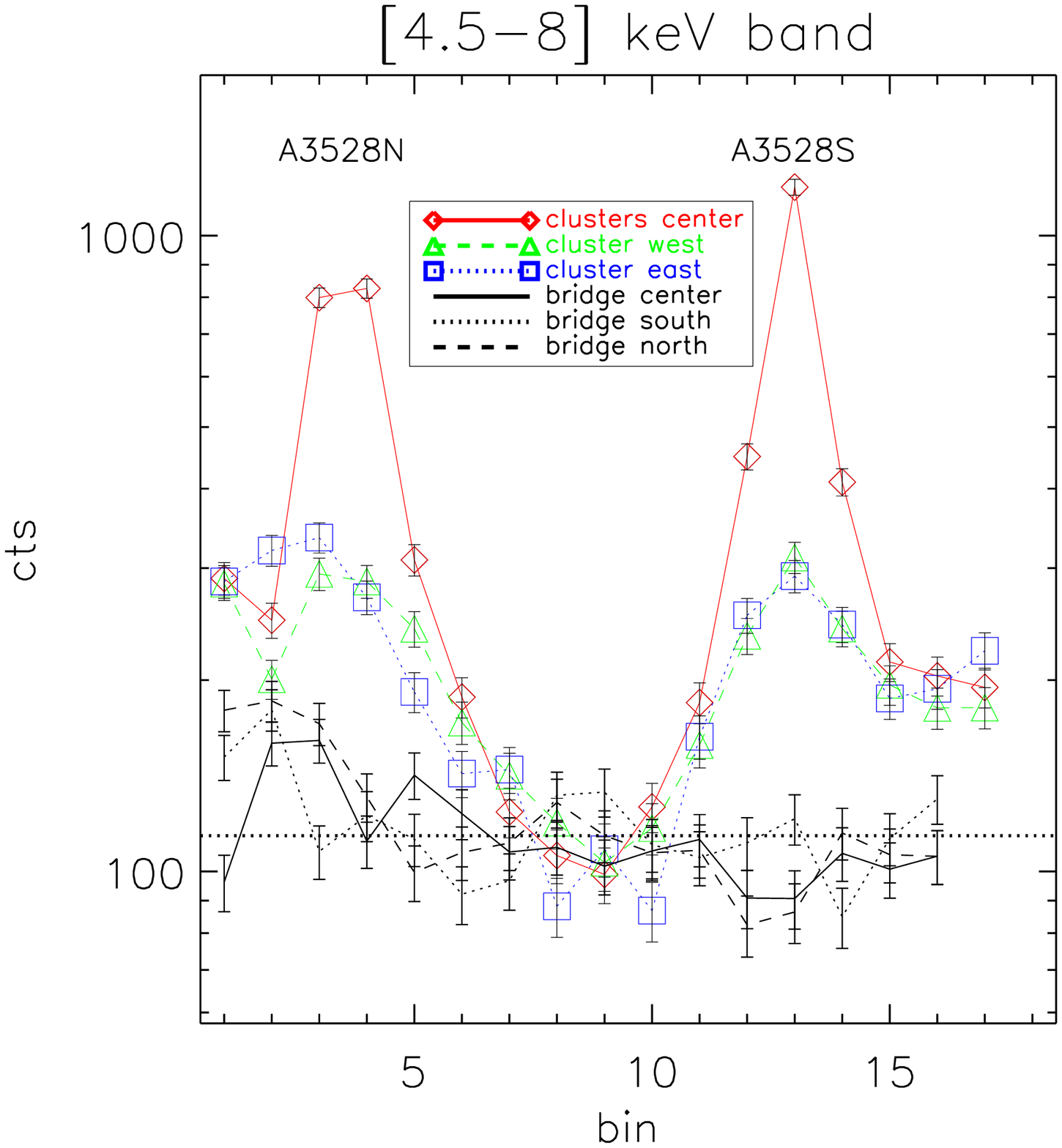,width=0.43\textwidth}
} 
\caption{{\bf (Left)} Adaptively smoothed exposure corrected EPIC image
of A3528 in the soft ([0.5-2.0] keV; {\it top panel}),
medium ([2.0-4.5] keV, {\it middle panel}) and hard band
([4.5-8.0] keV, {\it bottom panel}).
{\bf (Right)} Count rates in the regions defined in Fig.{\ref{clubridge}} and
obtained  in the soft ([0.5-2.0] keV; {\it top panel}),
medium ([2.0-4.5] keV; {\it middle panel}) and hard ([4.5-8.0] keV;
{\it bottom panel}) band. The error bars are the 1$\sigma$ Poisson
errors and the background shown as a dotted line is
obtained by a median of the counts in the two last bins West and East of the
three ``bridge'' regions.
Each bin has dimension  13.2$\times$10 pixels$^2$ 
($\approx 1.90\times 1.44$ arcmin$^2$; see caption of Fig.~\ref{clubridge}).
} \label{csmooth}
\end{center}
\end{figure*}

We perform (and cross-check) the vignetting correction in two independent 
ways: by applying it to (i) the spectra as is customary in the analysis
of EPIC data (Arnaud et al. 2001, Gastaldelo \& Molendi 2002) and 
(ii) the effective area via the use of the ancillary response file 
(arf) created through the SAS task \emph{arfgen}.

%%%%%%%%%%%%%%%%%%%%%%%%%%%%%
\subsection{Spatial analysis}
%%%%%%%%%%%%%%%%%%%%%%%%%%%%%

We use the clean linearized event files to generate MOS 1, MOS 2 and PN
images in the energy band [0.5-10] keV with a spatial binning of 8 arcseconds per pixel.
We merge together the three cameras images in order to increase the signal-to-noise ratio and the result is shown in Fig.{\ref{sources}}. 
In Fig.{\ref{optical}} we show the logarithmically spaced contour plots obtained
by the source cleaned EPIC image in the band [0.5-10.0] keV superposed 
on the optical DSS image. 
The two clumps, A3528-N and A3528-S, have a maximum at 
(RA, Dec, 2000)=$(12^{\rm h} 54^{\rm m} 22\fs1, -29^{\circ} 00' 46'')$ 
and $(12^{\rm h} 54^{\rm m} 40\fs6, -29^{\circ} 13' 44'')$,
respectively, with a comoving separation of 0.90 $h_{70}^{-1}$ Mpc
(a redshift of 0.053 is assumed for both the subclusters).

A diffuse faint emission appears to connect the two clumps.
In order to check the significance of this diffuse emission,
we perform an analysis of the counts along the regions depicted 
in Fig.{\ref{clubridge}}. We do this on EPIC images generated 
in three different energy bands, soft (0.5-2.0 keV), medium (2.0-4.5 keV) 
and hard (4.5-8.0 keV). The corresponding set of exposure maps 
for each camera has been prepared to account for spatial
quantum efficiency, mirror vignetting and field of view of each instrument
by running the SAS task \emph{eexmap}. The images from the three
instruments are then merged, weighting each of these by the ratio
of its maximum exposure time to the total maximum exposure time.
We restrict the higher bound of the hard energy band to avoid
effects from vignetting over-correction due to the prevalence
of the flat instrumental background in the broader energy band 4.5-10 keV.
We do not consider regions exposed less than 10\% of the total exposure.
From these exposure-corrected images, we obtain the surface brightness
profiles in the six regions marked in Fig.{\ref{clubridge}}.
These profiles are shown in Fig.{\ref{csmooth}} with the corresponding
adaptively smoothed, exposure corrected images in the three energy bands
adopted. The CIAO tool \emph{csmooth}, set to a minimum signal to 
noise ratio of 3, is used to produce these images.
The marked regions were choosen with the idea of looking for the enhancement of
counts along the ``bridge'' regions above the background. We use three regions
along the same directions to quantify the variations along the East-West and 
North-South
directions of the diffuse emission.  Obviously the
significance of plotting ``bridge'' and ``cluster'' regions together stands in
the overlap regions, where the bin have the same location.
A significant excess with respect to the background is detected
between the clumps in the soft and medium exposure-corrected images, while
on the contrary no excess is observed in the hard image.
In the hard band image, some edge effects cannot be avoided 
due to the low statistic and the level of background.

%%%%%%%%%%%%%%%%%%%%%%%%%%%%%%
\subsection{Spectral analysis}
%%%%%%%%%%%%%%%%%%%%%%%%%%%%%%

\begin{figure}[!]
\epsfig{file=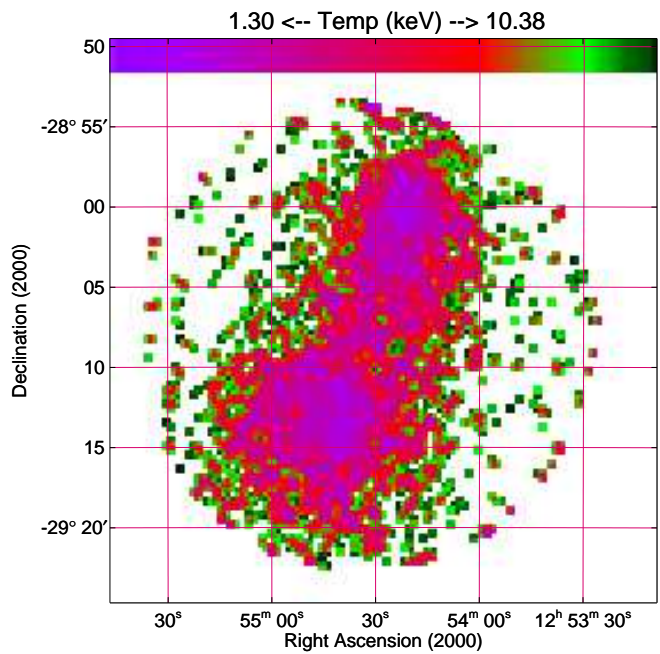,width=0.5\textwidth}
\caption{Temperature map obtained by using 5 X-ray colours
(0.4--0.8, 0.8--1.4, 1.4--2.0, 2.0--4.5, 4.5--8.0 keV)
and estimating the expected count rate with XSPEC for a
thermal MEKAL model, with fixed Galactic absorption \NH and metallicity.
The observed count rate is estimated in $4 \times 4$ original bins
of $(8.7 \times 8.7)$ arcsec$^2$ with step in X and Y of $2$ bins.  
}\label{fitcolor}
\end{figure}

We refer to Appendix A for details on the calibration
issues we deal with during our analysis. 
We focus here on the method used and the scientific results obtained.

The spectral analysis has been performed in successive steps.
From X-ray colours, we build a temperature map (see Fig.~\ref{fitcolor})
that we use to individuate interesting regions for further analysis.
The two clumps appear to have temperature between $2$ and $4$ keV
moving outward and do not show significant enhancement in temperature
in the region between them.
We then consider circles of 6 arcmins in radius centered on
each cluster's emission peak.
We use a MEKAL (Mewe et al. 1985; Liedahl et al. 1995) model in XSPEC 
(v.11.1.0, Arnaud 1996)
leaving as free parameters the column
density $\rm{N_{H}}$, the intracluster temperature $\rm{kT}$, 
the metallicity $\rm{Z}$ (in solar unity with respect to the values quoted
in Grevesse \& Sauval 1998), the redshift z and the
normalization. 
A3528 is at an optical redshift of 0.053 and the column density along the line of 
sight is $\rm{N_{H}} = 6.1\times10^{20}\,\rm{cm^{-2}}$. 
The results are shown in Tab.{\ref{tabfit}} in Appendix~B 
where in the upper half  we show the results obtained using
the effective area files, while in the bottom half the results obtained 
performing the vignetting correction directly on the events.
We confirm that the two methods agree well and adopt the one that
performs vignetting correction via the creation of an arf file
in the analysis that follows.

\begin{figure*}[!] 
\hbox{ 
 \hspace*{0.5cm} \epsfig{file=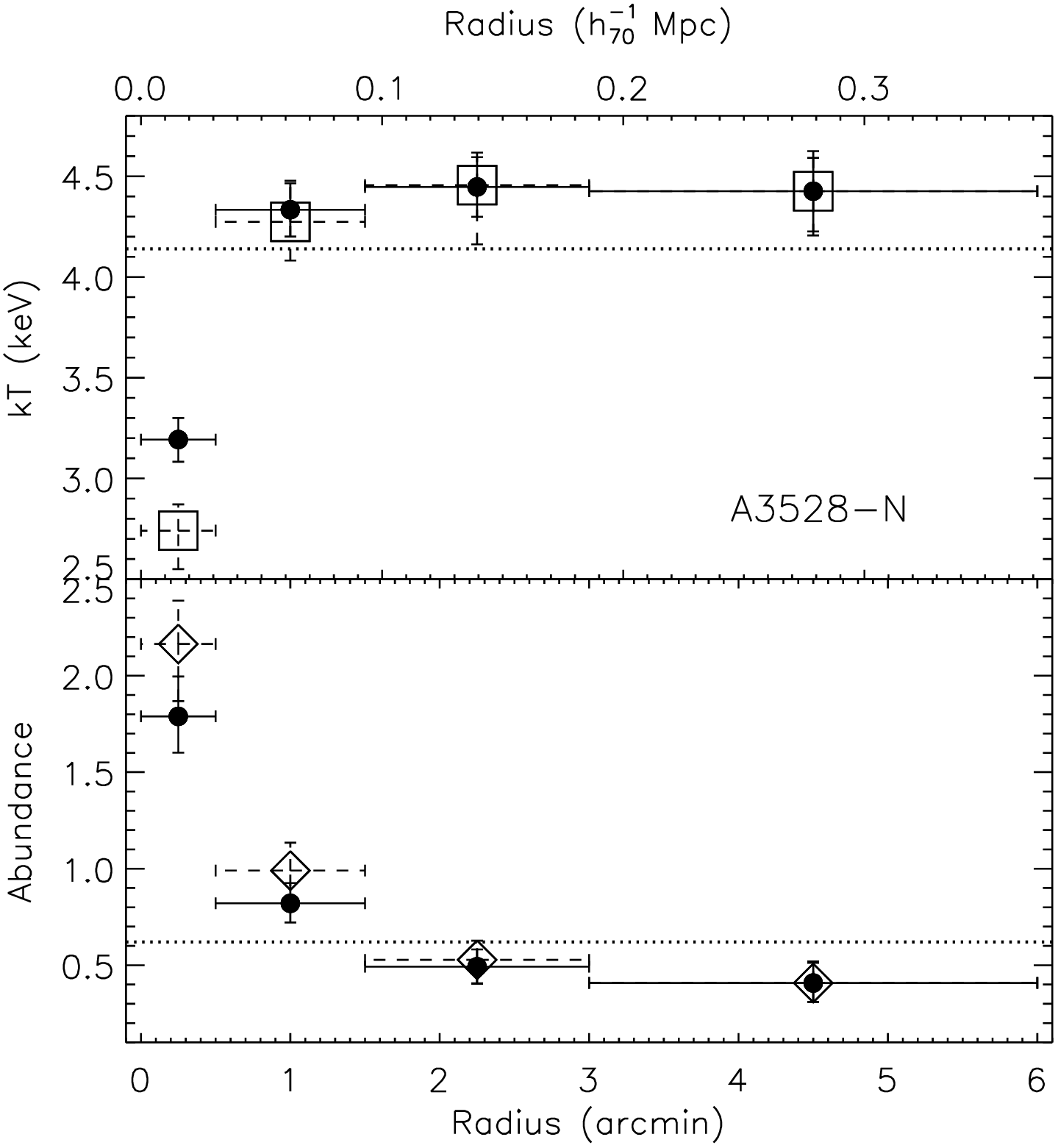,width=0.45\textwidth}
  \epsfig{file=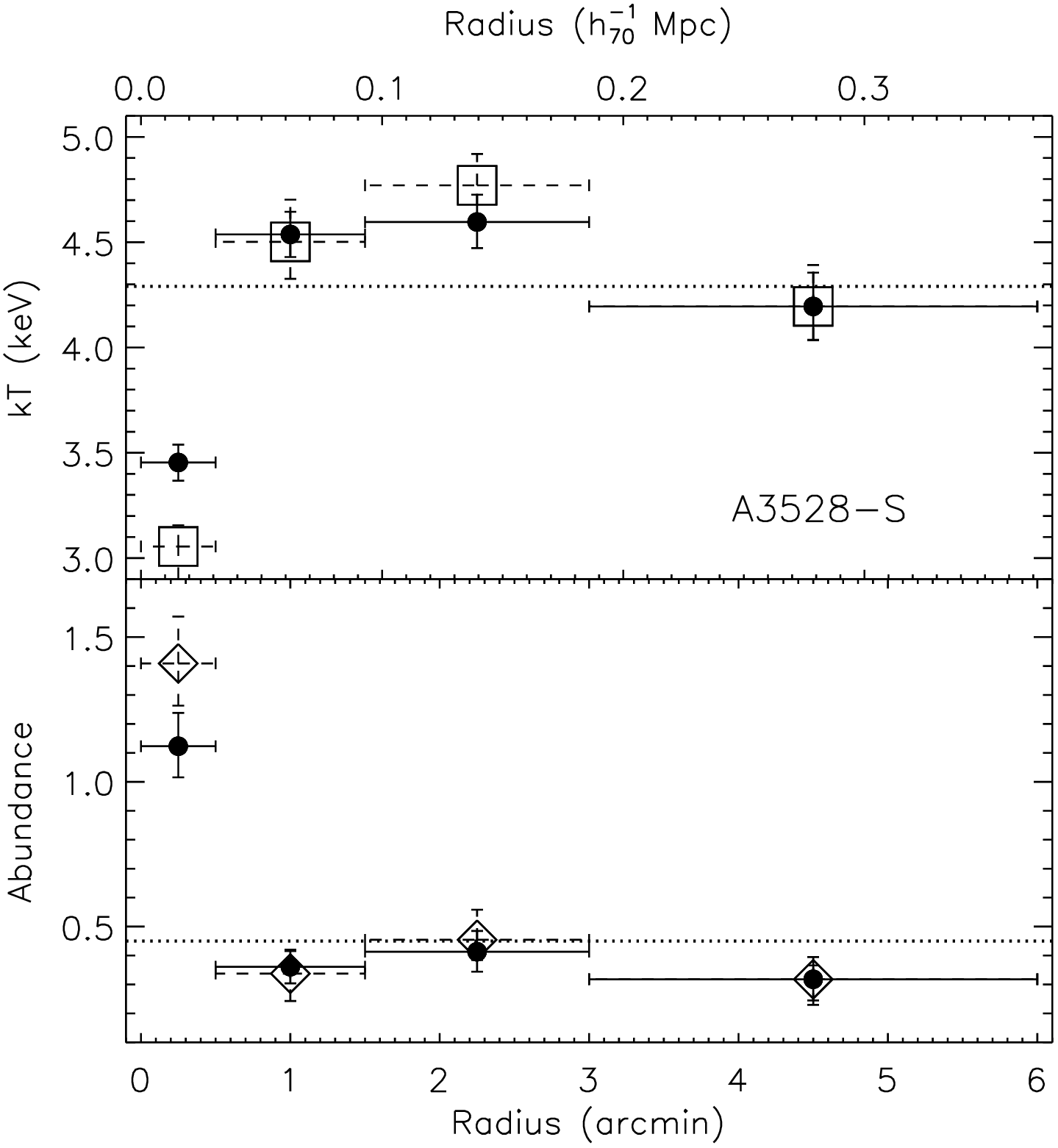,width=0.45\textwidth}
}
\caption{Temperature ({\it upper panels}) and metal abundance 
({\it lower panels}) profiles for A3528-N and A3528-S.
The {\it dots} show the best-fit results, while the {\it squares}
are the deprojected values.
The horizontal bars represent the width of the bins used to 
extract the counts and the dotted line indicates the temperature 
value from a spatially integrated fit. Uncertainties are at the 68\% level 
for one interesting parameter ($\Delta\chi^{2}=1$.)
}
\label{epictradial} 
\label{fig:prof}
\end{figure*}

\begin{figure*}[!]
\hbox{ \epsfig{file=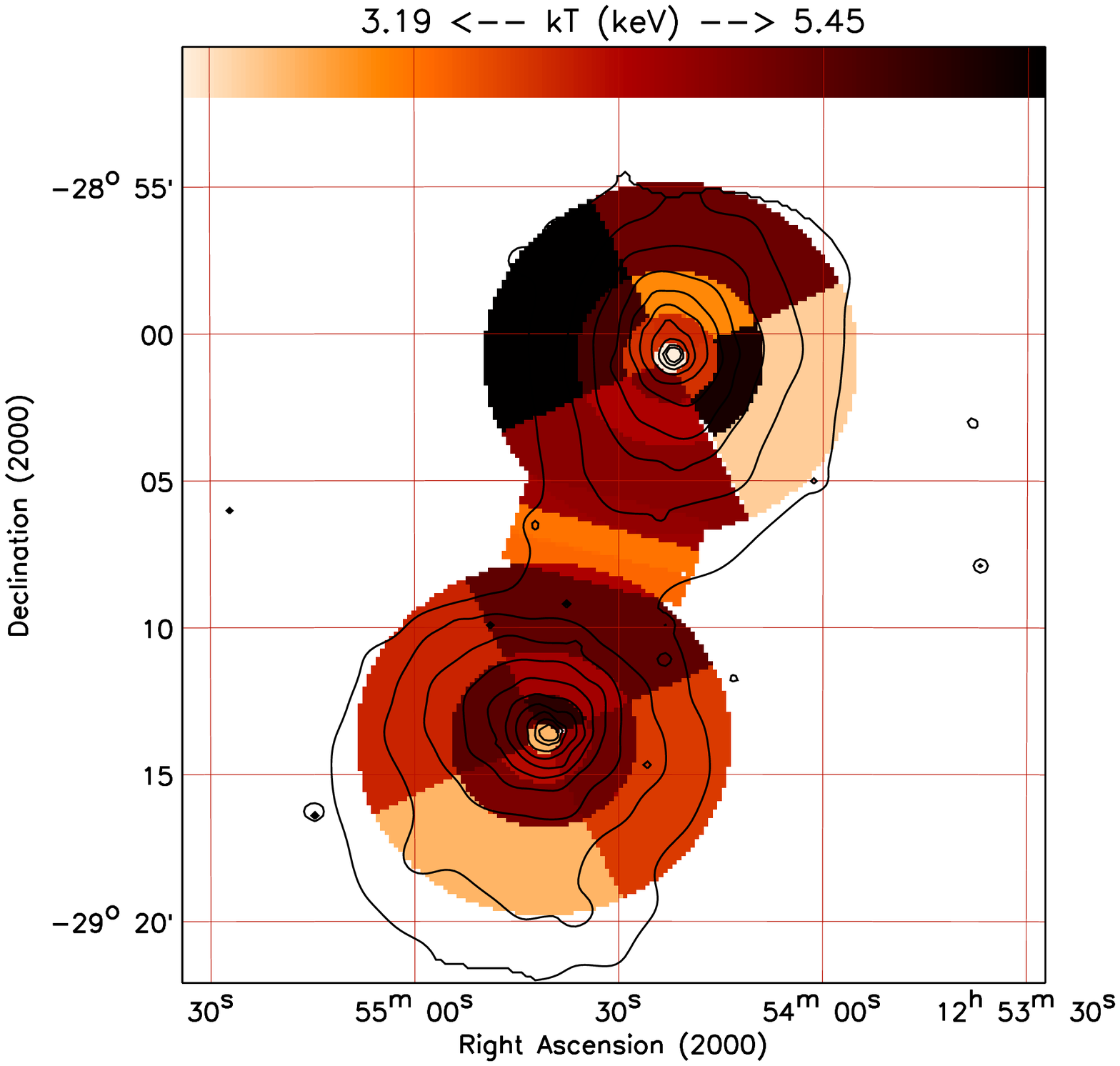,width=0.45\textwidth}
  \epsfig{file=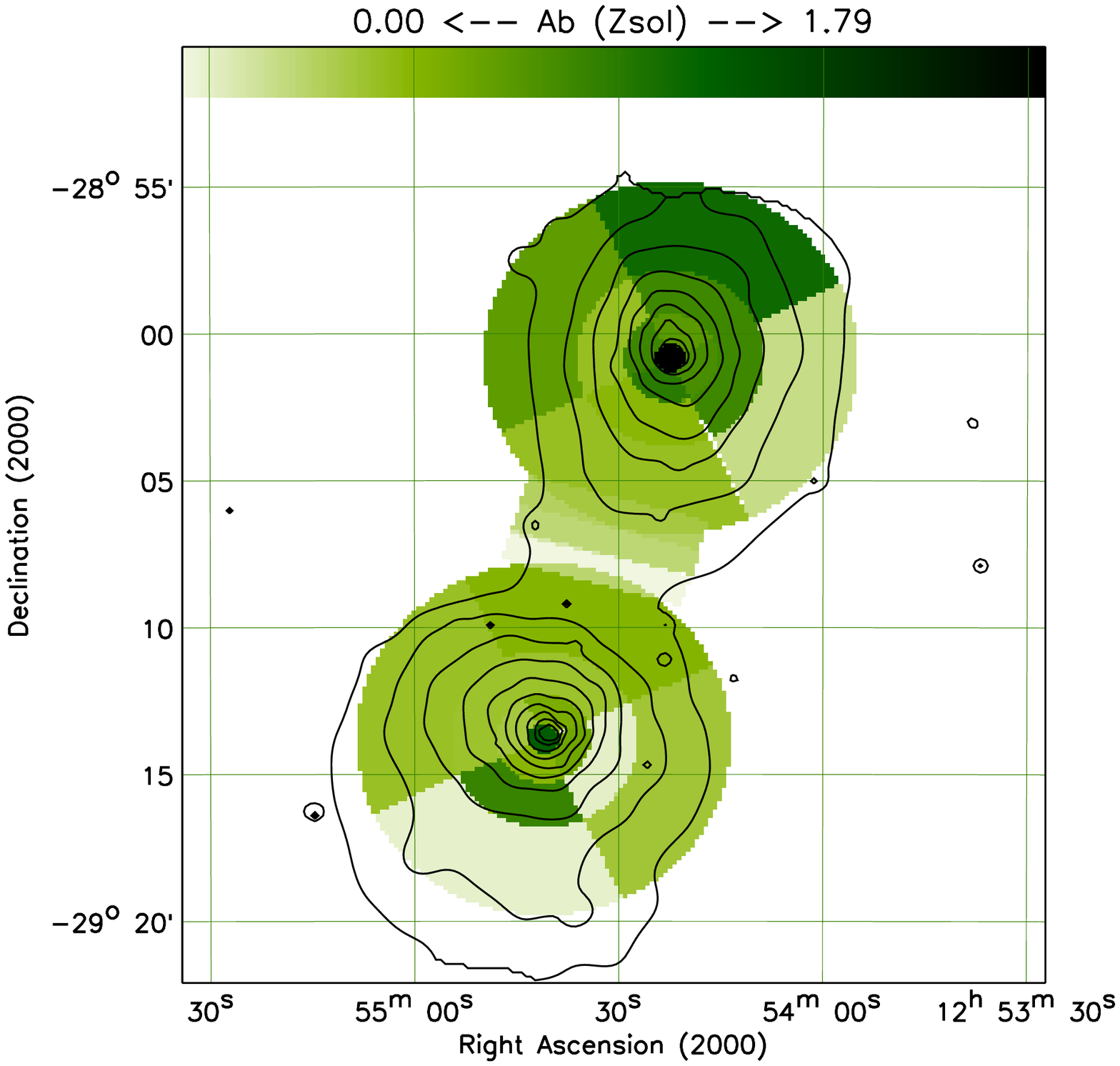,width=0.45\textwidth}
}
\caption{EPIC temperature map (left) and abundance map (right).
} \label{map}
\end{figure*}

\begin{figure*}[!]
\begin{center}
\hbox{ \epsfig{file=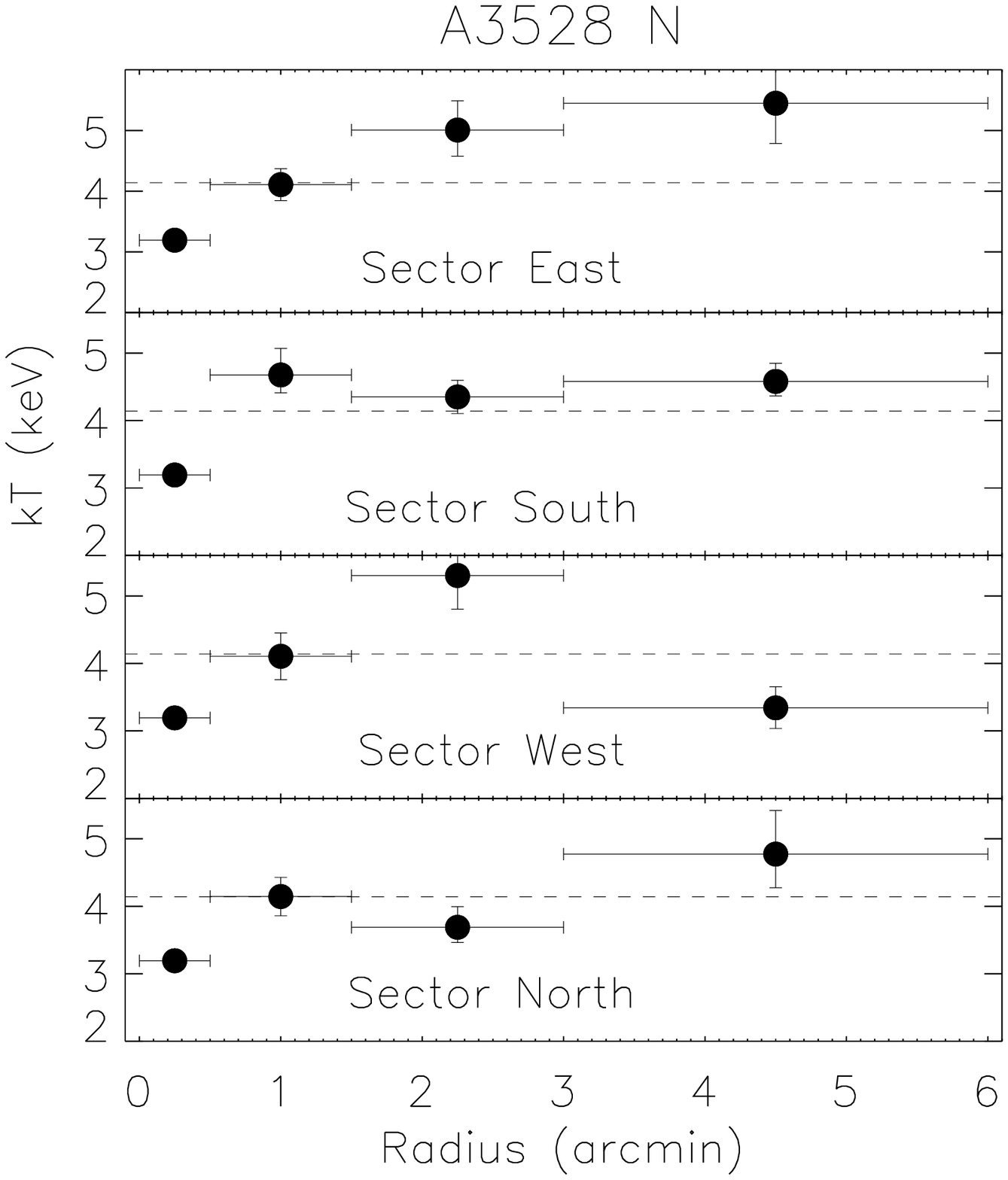,width=0.45\textwidth}
  \epsfig{file=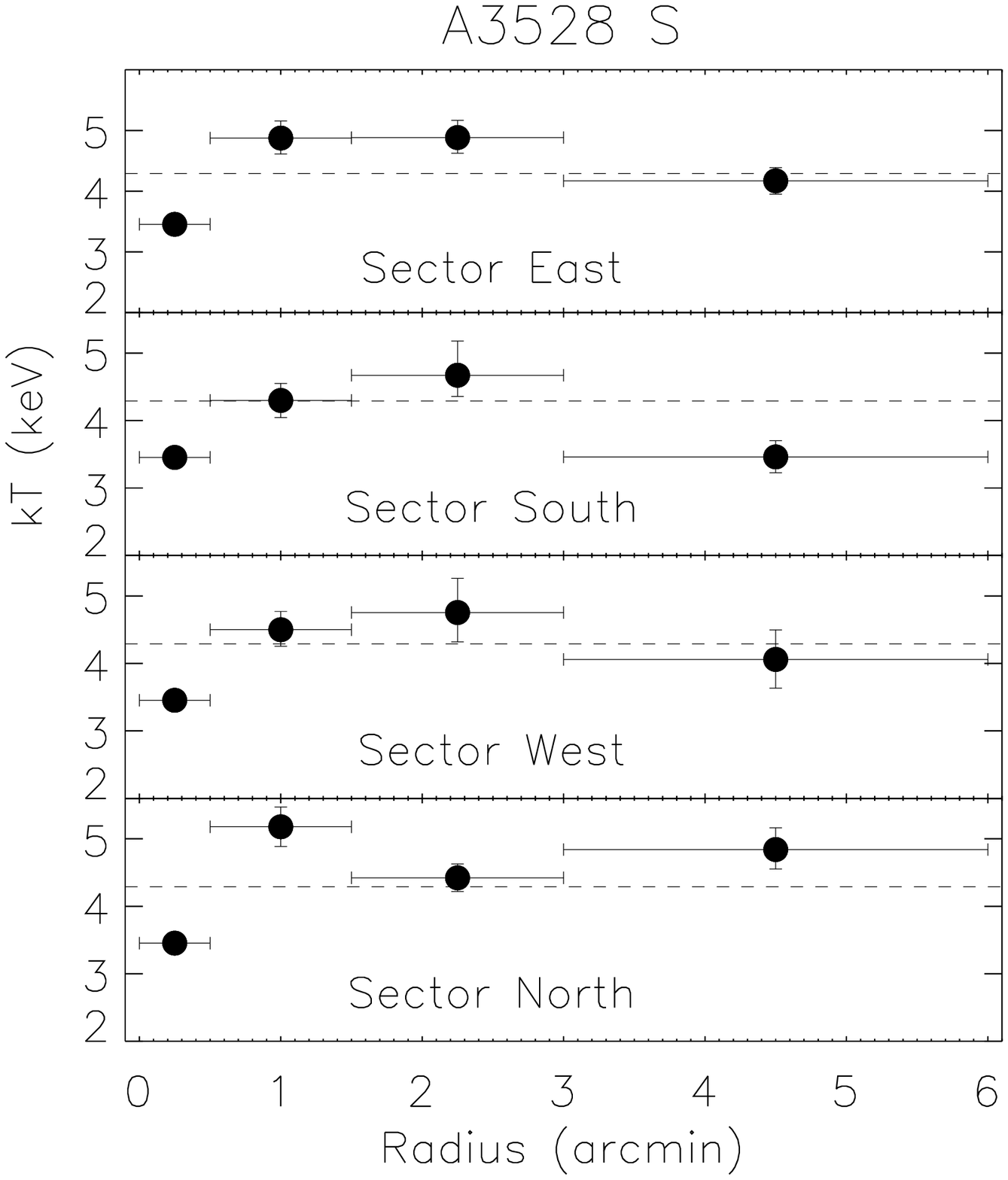,width=0.45\textwidth}
} \hbox{ \epsfig{file=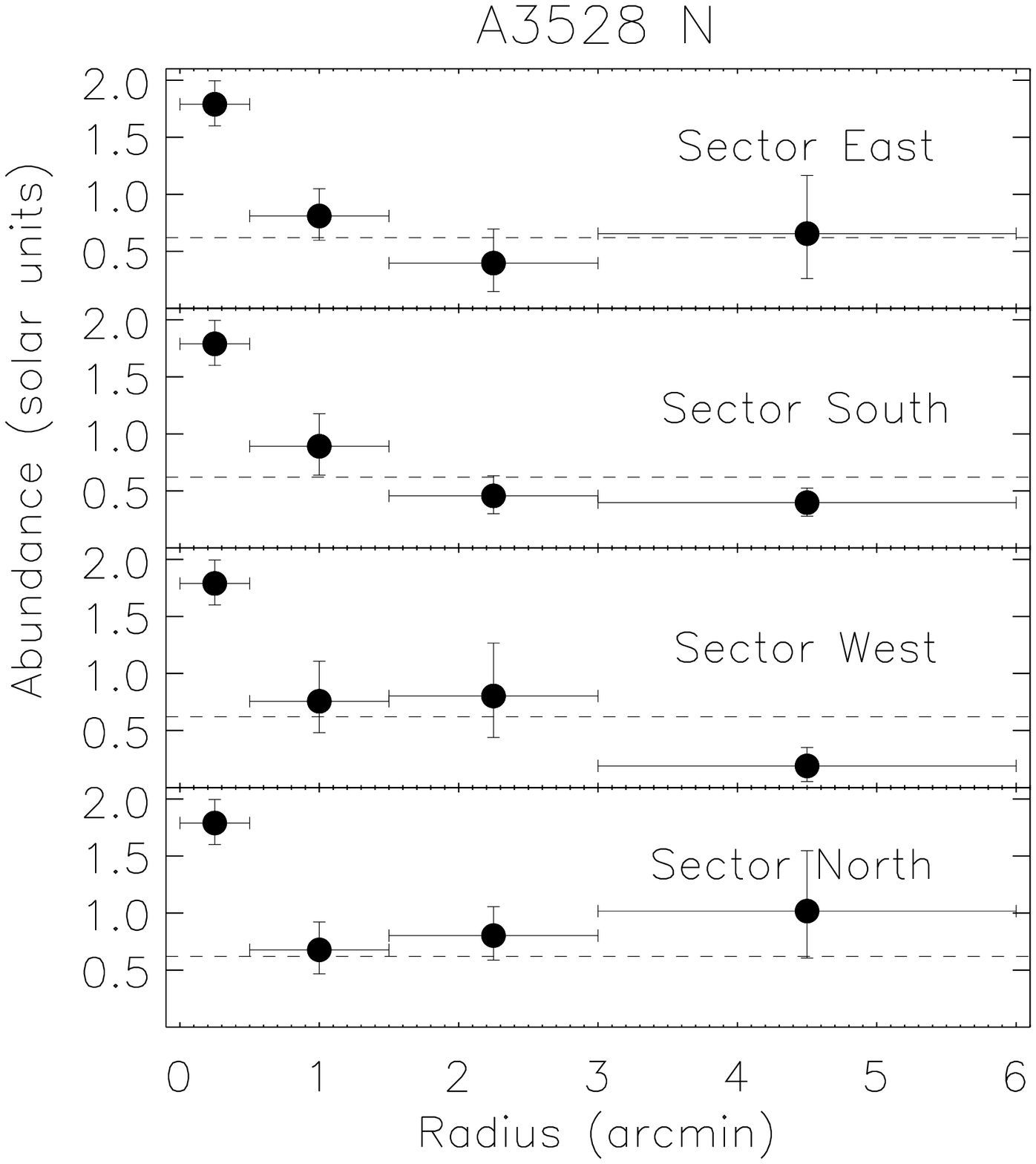,width=0.45\textwidth}
\epsfig{file=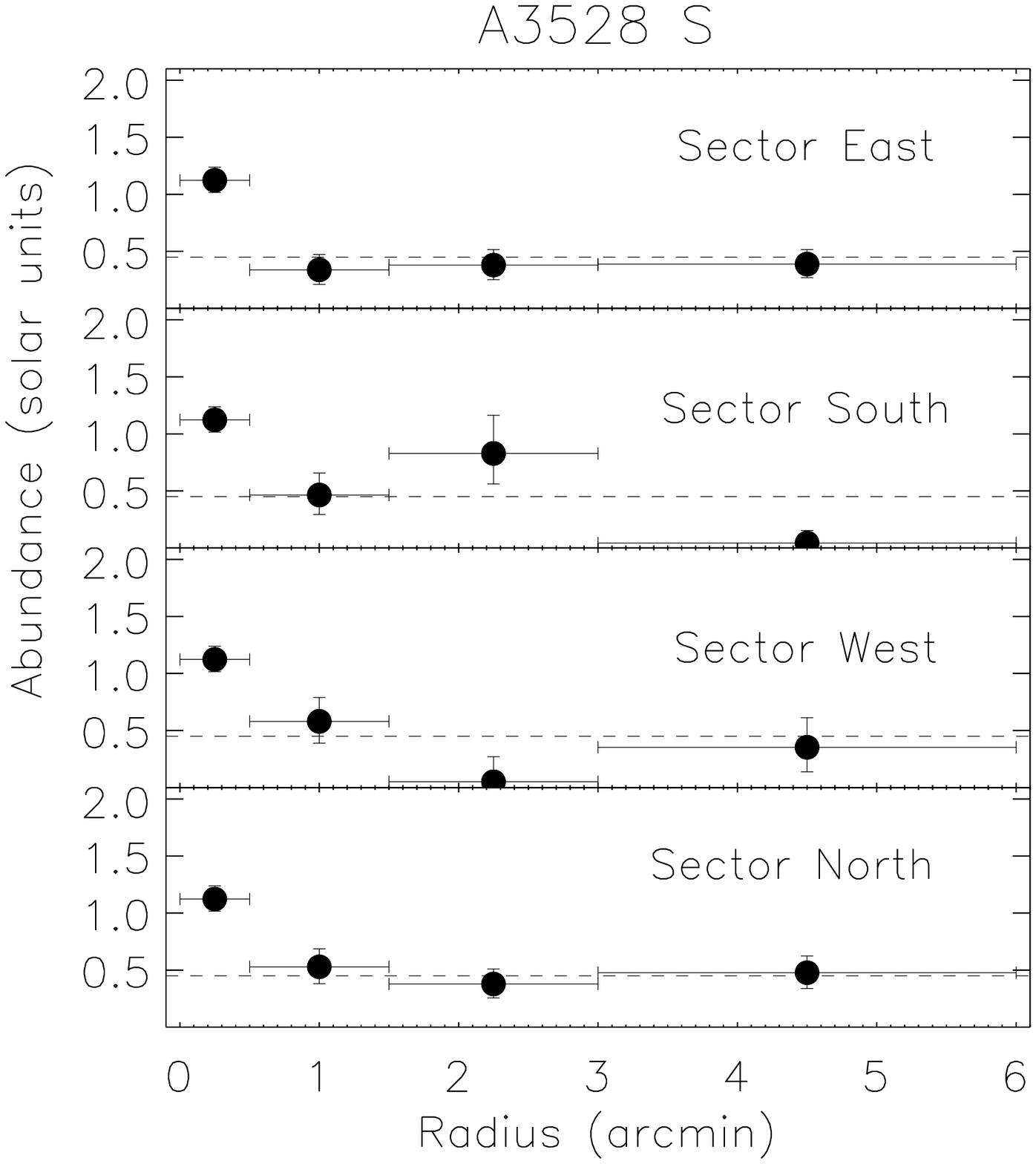,width=0.45\textwidth}
}
\caption{Gas temperature ({\it upper panels}) and metal abundance ({\it lower panels})
profiles measured in $90\deg-$sectors for A3528-N ({\it left}) and A3528-S ({\it right}). Uncertainties are at the 68\% level for one interesting parameter 
($\Delta\chi^{2}=1$). 
} \label{plotmap}
\end{center}
\end{figure*}

We try to correct for the uncertainties related to
calibration and background subtraction (see Appendix)
in a step-by-step procedure.
We fix the \NH at the Galactic value and the redshift at the optical 
determination and measure temperature and abundances in different
spectral bands (in function of the low energy cutoff, 
having found that different high energy cutoff, i.e. 8, 9 and 10 keV,
has no effects on the temperature determination)
and adopting different recipes 
for the background subtraction. We first use 
the plain background, in a second step we renormalize the background levels 
for different levels of intensity in the hard band ([8-12] keV) and in 
a last step we apply the double subtraction
method, described in Appendix A of Arnaud et al. (2002). 

For what concern the background, we are in condition to asses all 
its relevant components: the particle induced background 
is monitored in the hard part of the spectrum ($>$ 8 keV) and 
its count rate checked in order to renormalize the background 
field at the same count rate level;
moreover, the large area free of sources emission in our observation
allows to perform the double background subtraction and to take
control of the soft component of the cosmic X-ray background, 
that is estimated to be 75\% higher in the direction towards A3528
than towards our reference background field, APM 08279+5255, 
(see ROSAT survey maps through HEASARC X-ray background tool,
Snowden et al. 1997). 
To do that, we consider two circular regions of 3 arcmin-diameter to the 
West and East of A3528 and free of cluster emission, 
estimate the local background and compare it with what measured
in the corresponding regions of the blank field reprojected
on the same sky coordinates.
The count rate level in the hard X-ray
 band (8-12 keV) are: for MOS1 we have 
$2.06\pm0.12\times10^{-2}\,\rm{cts\,s^{-1}}$ for the source and 
$1.79\pm0.05\times10^{-2}\,\rm{cts\,s^{-1}}$ 
for the blank sky observation, for MOS2 
$1.98\pm0.12\times10^{-2}\,\rm{cts\,s^{-1}}$ for 
the source and $1.88\pm0.05\times10^{-2}\,\rm{cts\,s^{-1}}$ for 
the background  and for 
PN singles $5.86\pm0.31\times10^{-2}\,\rm{cts\,s^{-1}}$ and 
$5.22\pm0.09\times10^{-2}\,\rm{cts\,s^{-1}}$. 
We then use a normalization factor of 1.15 for MOS1, 1.05 for MOS2 and 1.12 for
the PN.

The results with different levels of background subtraction and different 
fitting energy bands are shown in Appendix C (cf. Fig.{\ref{appB_bands}}). 
By using the double subtraction method, which takes in principle
all the components of the background into account, and a joint fit 
of the three cameras, we measure an emission-weighted temperatures
within the inner 6 arcmin of $4.14\pm0.09$ keV for A3528-N and
$4.29\pm0.07$ keV for A3528-S (the fits are performed in the [0.5-8] keV 
band with galactic absorption \NH and redshift fixed).  
The overall abundances are $0.62\pm0.05$ $Z_{\odot}$for A3528-N 
and $0.45\pm0.04$ $Z_{\odot}$ for A3528-S
(reduced $\chi^{2}$ of 1.13 for 892 degrees of freedom and
1.05 for 1038 d.o.f. for A3528-N and A3528-S, respectively). 

We then analyze temperature and abundance profiles for the two subcluster
performing the same steps as for the global analysis.
The two subcluster emissions have been divided into four concentric annuli with
bounding radii $0'-0.5'$, $0.5'-1.5'$, $1.5'-3'$ and $3'-6'$.
We fit a single temperature model with fixed \NH and redshift to all
the spectra. We show in Figure~\ref{plotmap} the temperature and abundance
profiles obtained by using the double subtraction and a joint fit 
of the three cameras in the energy band 0.5-8.0 keV
(the profiles for each EPIC camera are shown in Appendix~C). 
All the spectra can be properly fitted with a single temperature model,
including the innermost bins ($\chi^{2}=210$ for 191 d.o.f. and
$\chi^{2}=298$ for 287 d.o.f. for the central bins of A3528-N and 
A3528-S, respectively).
The fits are not improved by adding another thermal component.

The two subclusters show clear evidence of a cool core, with a 
peaked surface brightness and a steep gradient in metal abundance,
the latter being more strong in A3528-N (see Fig.~\ref{fig:prof}). 
An estimate of the cooling time, $t_{\rm cool} \approx
(3/2) n_{\rm gas} T_{\rm gas} \times Vol / L_{\rm X, bol}$,
in the inner 30 $h_{70}^{-1}$ kpc of 
A3528-N and A3528-S gives values of 0.8 and 1.2 Gyr, respectively,
an order of magnitude lower than the expected age of the
Universe at $z=0.053$ (12.7 Gyr).

With XMM we can provide a detailed temperature map and in particular study
in greater detail
the northern subcluster and directly the region between the two subclusters, 
which were affected by the ROSAT supporting structure 
(see Fig.8 of Reid et al. (1998)).
We fit the EPIC cameras together since the beginning, in 
order to increase the signal to noise ratio, fixing the band at [0.5-8.0] keV
and with \NH and redshift fixed. 
We divide each annular region analyzed before in four angular sectors 
90 degrees wide  and in addition we analyze three central boxes of 
$6 \times 1$ arcmin in the region between the two subclusters, to
address the presence of any increase in the temperature due to shocks. 
In all the regions used we have more than 70\% of source counts.  
The results are shown in EPIC temperature and abundance maps 
in Fig.{\ref{map}}.
In Fig.{\ref{plotmap}},  the temperature and abundance 
profiles in the four sectors are shown for A3528-N and A3528-S.
The regions between the two clusters show no evidence of temperature
enhancement as the temperature and abundance values in the three 
central boxes are, going from North to South, 
$4.46^{+0.61}_{-0.50}$ keV, $3.90^{+0.66}_{-0.60}$ keV, $3.79^{+0.53}_{-0.52}$
 keV. The abundances have large errors due to low statistic in the iron line 
and are $0.27^{+0.37}_{-0.27}$, $0.14^{+0.27}_{-0.14}$, $0.0^{+0.26}_{-0.0}$
$Z_{\odot}$.

%%%%%%%%%%%%%%%%%%%%%%%%%%%%%%%%%%%%%%%%%%%%%%%%%%%
\section{The state of merging at other wavelengths}
%%%%%%%%%%%%%%%%%%%%%%%%%%%%%%%%%%%%%%%%%%%%%%%%%%%

%%%%%%%%%%%%%%%%%%%%%%%%%%%%%%%
\subsection{In the optical}
%%%%%%%%%%%%%%%%%%%%%%%%%%%%%%%

\begin{figure}[!]
\hspace{-0.5truecm}
\epsfig{figure=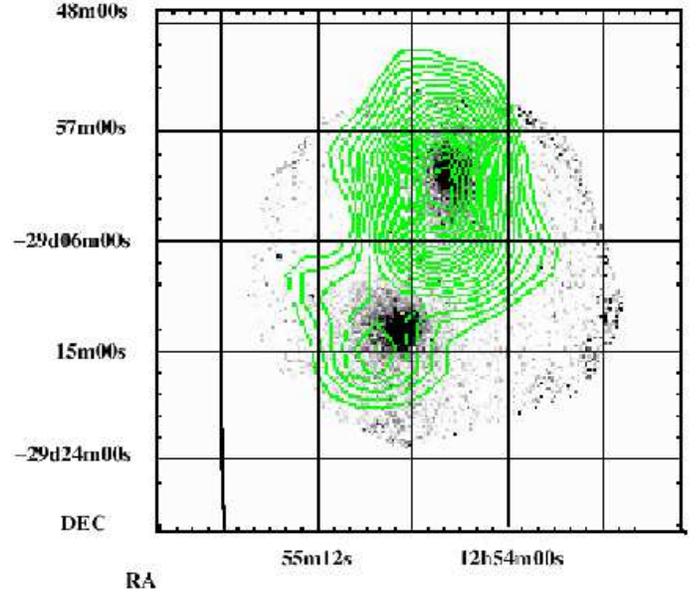,width=0.6\textwidth}
\caption{Isodensity distribution of the light in the A3528-N and A3528-S 
region, from the COSMOS/UKSTJ galaxy catalogue, over plotted onto the X-ray 
image. These contours
have been calculated by summing the galaxy luminosities in bins
of $2\times 2$ arcmin and then smoothed with a Gaussian of 3 pixels FWHM.
} \label{optcontour} \end{figure}

In the X-ray band, A3528-N and A3528-S appear to be very similar both
in luminosity and in mass. On the contrary, considering the distribution 
of optical galaxies, there is an evident asymmetry between the two 
subclumps, being A3528-N clearly dominant with respect to A3528-S.

In Fig.{\ref{optcontour}} we show the isodensity contours of the galaxy light 
from the COSMOS/UKSTJ galaxy catalogue (Yentis et al. 1992), 
overplotted onto the X-ray image. These contours have been derived
by summing the luminosities of galaxies in bins of $2\times 2$ arcmin 
and then smoothed with a Gaussian of 3 pixels FWHM.
This procedure gives more information with respect to a simple number 
weighted binning, because it takes into account the possibility of
a difference in the luminosity distribution of the two clusters. 

From this figure it is immediately clear the difference in total 
luminosity of the two subclumps, being A3528-N about 8.6 times more
luminous than A3528-S.
Moreover, while there is a good agreement between the position 
of the peak of the light distribution, the cD galaxy and the X-ray peak
in A3528-N, the maximum in the light map of A3528-S
is shifted by $\sim 2.8\,\rm{arcmin}$ to the South 
with respect to the position of the cD galaxy (that coincides 
with the X-ray peak; see, e.g., Fig.~\ref{optical}).
In the case that a merging is taking place along the North-South
direction, this shift is in opposition to what expected, 
with the galaxies that tend to preceed the hot gas 
(see, e.g., Tormen, Moscardini \& Yoshida 2003).

It is quite difficult to explain why these two clumps with very similar 
X-ray properties appear so different in the optical band. 
This fact, added to the relative shift between light and hot gas distribution, 
seems to indicate that an interaction between the more external, 
non-collisional part of the clusters (the galaxies) is already in place, 
leading to a loss of optical light in A3528-S. 
This could be an off-axis merging 
where the two clusters are orbiting around each other: however, this scenario 
does not explain why A3528-N, although having the same mass, appears more 
relaxed.
The reason could be a different dynamical history of the two clumps, i.e. 
A3528-S suffered more than A3528-N of tidal forces during its travel 
in the A3528 complex.

Unfortunately, no help comes from the velocities, being impossible to 
divide the contribution of the two clumps (Bardelli et al. 2001):
there is only an indication of two peaks in the velocity histograms, 
corresponding to the velocities of the two cD galaxies (16564 km s$^{-1}$ 
for cD North and 16923 km s$^{-1}$ for cD South) but the overall distribution 
is significantly consistent with a single Gaussian of 
$\sigma \sim 888$ km s$^{-1}$.

%%%%%%%%%%%%%%%%%%%%%%%%%%%%%
\subsection{In the radio}
%%%%%%%%%%%%%%%%%%%%%%%%%%%%%

The radio properties of A3528-N and A3528-S do not provide 
strong constraints about the status of the merging in A3528,
nevertheless some considerations can be made. 
The radio luminosity function (RLF) for early--type galaxies,
which reflects the probability of an early--type galaxy of
given optical magnitude to develop a radio galaxy above a given
radio power, is in agreement with the "universal" RLF 
for early--type galaxies (Venturi et al. 2001). Even though there 
is not yet a general consensus on the effect of cluster mergers 
on the RLF (see for instance Miller \& Owen 2003
and Venturi et al. 2002), it is most likely that the RLF in the A3528 
complex reflects a virialized situation.

Another striking feature of the radio emission in A3528 is the
presence of 5 extended radio galaxies. Three of them are narrow
angle tail (NAT) sources and are located at the centre of A3528-N,
at the centre of A3528-S and in between them, respectively.
In all five cases, multifrequency high resolution radio observations 
show an active nucleus and twisted/distorted jets and lobes (Venturi 
et al. in prep). The radio emission associated with the dominant galaxies of 
A3528-N and A3528-S  is all embedded in the optical light, while in the 
other cases the projected linear size exceeds that of the optical 
counterpart and limited to $\sim$ 30 -- 45 kpc. 
It is noteworthy that the tails of the two centrally located NATs 
point in the same direction (South--West to North--East). It is now 
believed that a combination of ram pressure (due to the galaxy velocity
with respect to the intracluster medium) and "cluster weather" 
(i.e. bulk motion in the intracluster gas) are the mechanisms responsible 
for the tail bending (see Feretti \& Venturi 2002 for a brief and recent 
review). Considering that the tails are very short and cluster weather
may not be so efficient on such small scales, it is likely that ram pressure 
is here the dominant mechanism for the formation of the two NATs.

Finally, the lack of indication of extended cluster scale radio 
emission also deserves a comment. It is now accepted that cluster
mergers can provide the necessary electron reacceleration to
produce radio halos (see for instance Buote 2002; Giovannini \& Feretti
2002; Sarazin 2002 for recent reviews), provided that enough relativistic
electrons are deposited in the intracluster medium. Cluster active radio 
galaxies are the most likely sources of electron replenishment. 
The high number of active radio galaxies in A3528-N and A3528-S indicates
that the presence of relativistic particles is not an issue here, however
there is no indication of a radio halo. This may suggest that particle
acceleration induced by merger, if present at all in the central regions
of two clumps, is not efficient to reaccelerate the electrons at the 
levels requested to form a radio halo.
None of these pieces of evidence is conclusive, however they all
seem to suggest that a major merger (or an on--axis merger) between the 
two structures has not (yet) taken place.

%%%%%%%%%%%%%%%%%%%%
\section{Discussion}
%%%%%%%%%%%%%%%%%%%%

The two clumps that compound A3528 have the distinctive features of
relaxed structures, with a centrally peaked surface brightness
(see Fig.~\ref{sources}), a cool core and a steep gradient
in metal abundance (see Figs.~\ref{fig:prof} and \ref{plotmap}).
A X-ray soft, bridge-like emission (as shown in Fig.~\ref{csmooth})
connects them.
Considering also the absence of characteristic enhancement
in the gas temperature both in the X-ray colours (Fig.~\ref{fitcolor}) 
and spectral (Fig.~\ref{map}) maps and sharp 
discontinuities in the surface brightness distribution,
we have no indication of shock heated gas and 
that a merging action is now in progress, at least a major head-on
(i.e. impact parameter equals to zero) merger.

\begin{figure}
\hspace{0.5truecm}
\epsfig{figure=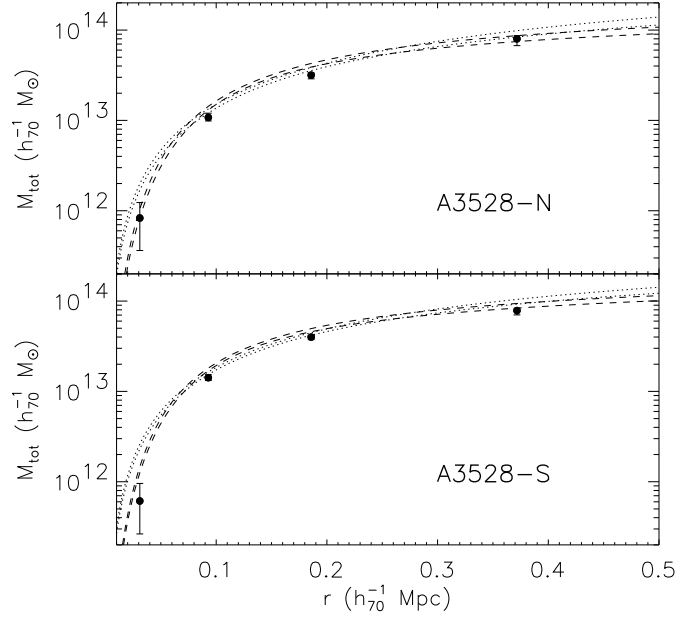,width=0.5\textwidth}
\caption{
Mass profiles of A3528-N and A3528-S
({\it points}: from hydrostatic equation applied to the spectral deprojected
results; {\it dashed lines}: $1 \sigma$ range values obtained by assuming
a King profile; {\it dotted line} $1 \sigma$ range values obtained
by assuming a NFW profile).
} \label{fig:mass} \end{figure}

To assess the dynamics of the two subclumps forming A3528, we recover
the gravitational mass profiles
in these systems by using the deprojected best-fit spectral results
obtained for azimuthally averaged annular spectra
(see Fig.~\ref{fig:prof} and details in Ettori, De Grandi, Molendi 2002).
Assuming the spherically symmetric distribution of the
X-ray emitting plasma in hydrostatic equilibrium with the
underlying dark matter potential, we infer a total
gravitating mass of $8.0^{+0.7}_{-1.3} \times 10^{13} h_{70}^{-1} M_{\odot}$
and $7.8^{+0.5}_{-0.8} \times 10^{13} h_{70}^{-1} M_{\odot}$ for
A3528-N and A3528-S, respectively, within 6 arcmin ($0.38 h_{70}^{-1}$ Mpc).
We also assume two functional forms of the dark matter
potential,  the King approximation to the isothermal sphere
(Binney and Tremaine 1987) and
a Navarro, Frenk \& White (NFW) (1997) profile,
\begin{eqnarray}
\lefteqn{M_{\rm tot, model}(<r) = 4 \pi \ r_{\rm s}^3 \ \rho_{\rm s} \ f(x), }
\nonumber \\
\lefteqn{ \rho_{\rm s} =  \rho_{\rm c} \frac{200}{3} \frac{c^3}
{\ln (1+c) -c/(1+c)} },  \\
f(x) & = &  \left\{ \begin{array}{l}
\ln (x+\sqrt{1+x^2}) - \frac{x}{\sqrt{1+x^2}} \; {\rm (King)} \nonumber \\ 
\ln (1+x) - \frac{x}{1+x} \ \hspace*{1cm} {\rm (NFW)} 
\end{array}
\right.
\label{eq:mass_nfw}
\end{eqnarray}
where $x = r/r_{\rm s}$, $\rho_c$ is the critical density and
the relation $r_{\Delta=200} = c \times r_{\rm s}$ holds for the
NFW profile.
We try to assess
which is the mass model that better reproduces the deprojected gas temperature
profile by inverting the equation of the hydrostatic equilibrium.
We obtain that the King profile is a better modelling of the data
($\chi^2 = 3.9$ and $7.0$ for A3528-N and A3528-S, respectively,
whereas a NFW profile gives $7.9$ and $14.9$ for 2 degrees-of-freedom)
with best-fit parameters $(r_{\rm s}, c) = (86 \pm 6$ kpc, $11.2 \pm 0.6)$
and $(75 \pm 4$ kpc, $13.1 \pm 0.5)$ for A3528-N and A3528-S, respectively
[$(265 \pm 47$ kpc, $5.2 \pm 0.6)$ and $(197 \pm 26$ kpc, $6.8 \pm 0.6)$
for a NFW profile; see Fig.~\ref{fig:mass}].
From the best-fit results of the King profile, we can extrapolate
the mass expected in these clumps at an overdensity of 200 with respect
to the critical density, and we obtain a total virial mass of
$1.5 \times 10^{14} h_{70}^{-1} M_{\odot}$ and
$1.6 \times 10^{14} h_{70}^{-1} M_{\odot}$ for A3528-N and A3528-S,
respectively, at about $r_{200} = 1.1 h_{70}^{-1}$ Mpc.
In a simple, but still extreme, scenario in which the two systems
lie on the plane of the sky, with the apparent distance between
the centers of 0.9 $h_{70}^{-1}$ Mpc being their 3D spatial separation,
and are falling from infinity with zero angular momentum and
zero initial velocity toward each other, their relative velocity,
$v = \sqrt{2 G (M_1 + M_2)/ R}$,
should be of $\sim$1720 km s$^{-1}$ and their centers should cross in
about 0.5 Gyr. This infall velocity should imply a Mach number
of $M \sim 2$ that should induce a visible shock in this head-on 
configuration.

To estimate what the presence of a shock would produce and whether we are in 
condition to record it, we assume an adiabatic shock where all the 
dissipated shock energy is thermalized
and apply the standard Rankine-Hugoniout jump conditions
(Landau \& Lifshitz 1959, Sarazin 2002). If we further assume a realistic 
Mach number of about $2$ (the motions in cluster mergers are expected to be
moderately supersonic with Mach numbers $M \lesssim 3$, Sarazin 2002), we 
should expect a compression factor $C\approx2.3$ and  a gas hotter by more 
than 70\% should be measured on scales 
of 1 arcmin or so (see, e.g., Figure~7 in Ricker \& Sarazin 2001, 
where shock heated gas is observed on this scale in hydrodynamics 
simulations of colliding clumps with impact parameter of 0 and 0.5 
Gyr before the maximum luminosity reached at the crossing of the
cores). 
If the gas is at the ambient temperature  
of 4 keV, we in-fact expect to see gas of a factor $2.3$ 
more dense at temperatures of $\sim 7$ keV.
None of these hints of the presence of a shock 
are detected in the surface brightness 
distribution and in the temperature map.
This evidence is, in some way, surprising considering the relative
small separation between the clumps and their large masses.
For example, a similar structure is observed in Cygnus A 
(Markevitch et al. 1999), where two clumps, at $T_{\rm X} \sim 4-5$ keV 
and at a distance of 1 Mpc,
collide and form a well defined shock between them that  
significantly affects the ASCA temperature map.

\begin{figure}[!]
\epsfig{figure=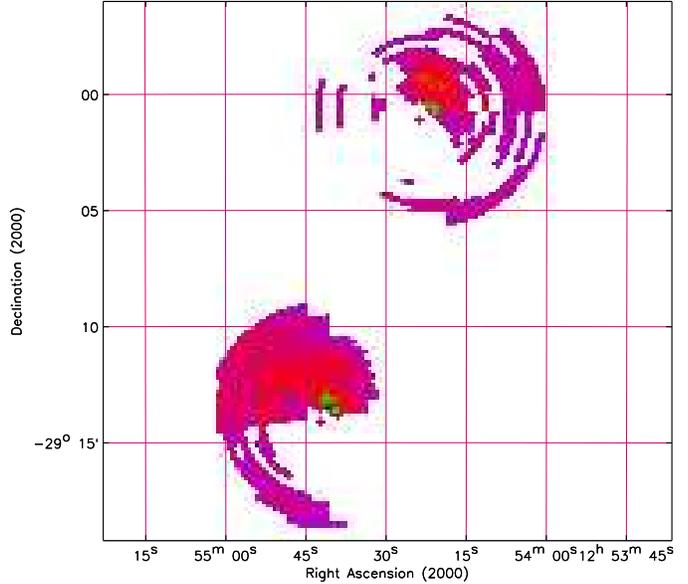,width=0.5\textwidth}
\caption{
Residuals above $2$ in the X-ray brightness
distribution. Given two regions, $A$ and $B$, of (length, width)${}
= (2$ pixels$\approx 17$ arcsec$, 30^\circ)$ located symmetrically with
respect to the center and with observed $C_A$ and $C_B$ counts respectively,
the residuals are estimated as $\sigma = (C_A - C_B) / \sqrt{C_A + C_B}$.
An azimuthal scan with step of $30^\circ/3$ was done to smooth the map.
} \label{asim} \end{figure}

So far, we have adopted the working hypothesis, also supported
from recent literature (e.g. Bardelli et al. 2001, Donnelly et al. 2001), 
that the two subclumps in A3528 are involved in a pre-merger along
the North--South direction with zero impact parameter.
This hypothesis fails, however, to explain
the optical appearance and has remarkable difficulties to
motivate the X-ray appearance of A3528, lacking any feature
that characterize a merger between these two massive subclusters
at a relative so small separation (and again, it is stringent
the difference with the two clumps of Cygnus A which are separated by
about 1 Mpc).

We consider here another explanation for what is observed, i.e. that the
two subclumps are in an off-axis, post-merger phase.
The X-ray emitting gas appears firmly confined in the central 
deep potential well of the two systems, giving rise to well 
defined gradients in surface brightness, temperature and metallicity 
profiles, whereas it seems more disturbed in the outskirts.
As we show in Fig.~\ref{asim}, 
the surface brightness is azimuthally asymmetric with significant 
excesses in emission (with respect to the counter-part in 
opposition to the X-ray center) in the North--West region of A3528-N 
and in the North--East area of A3528-S. The latter X-ray excess
coincides spatially with an enhancement in the optical light
distribution.
On the other hand, as we discuss in Section~4.1,
no segregation in two clumps is observed in the galaxy distribution, 
being A3528-S marginally detectable (and the X-ray map confirms it
as the most disturbed system).
These results point toward a scenario in which the A3528-N and 
A3528-S are moving along the N-W/S-E and N-E/S-W direction, respectively,
with a merger that had the closest encounter of the cores
about 1--2 Gyrs ago with an impact parameter $b\approx 5 r_{\rm s}$
and that was sufficient to disturb the outer galaxy distribution
but not to affect the cluster cores themselves
(Figure~4 in Ricker \& Sarazin (2001) provides 
a reasonable representation of the kind of merger here suggested).
We also note that this impact parameter is consistent with the observed 
separation between the clumps and the measured scale radius $r_{\rm s}$
in the NFW profile.   
Moreover, and following the results in Ricker \& Sarazin (2001),
this picture could explain the narrowness of the bridge connecting
the two subclumps, whereas we should have expected a more extended emission
in a head-on collision.
Finally, the late stage of the merger and the non-zero impact parameter
justify the absence of any detectable shock and, in particular, 
of a $M \sim 2$ case, estimated above from the calculated
free-fall velocity, which holds only for head-on mergings 
(and this could be another point of difference in the
comparison between A3528 and Cygnus~A, where Markevitch and collaborators 
found good agreement between the velocities derived from the temperature 
jumps and the values predicted by free fall, corroborating the head-on merger
hypothesis).

%%%%%%%%%%%%%%%%%
\section{Summary}
%%%%%%%%%%%%%%%%%

Using the high spatial resolution and large effective area of the 
EPIC instruments on board of \xmm, we have performed a spatial analysis 
of the surface brightness, 
gas temperature and metal abundance distributions of the 
double cluster A3528 in the Shapley supercluster.

The main conclusions of our work are:
\begin{itemize}
\item the two clumps, A3528-N and A3528-S, have a mean gas temperature 
of $4.14 \pm 0.09$ keV and $4.29 \pm 0.07$ keV, respectively,
that significantly decreases in the inner 30 arcsec (31 $h_{70}^{-1}$
kpc) to a deprojected value of $2.75^{+0.12}_{-0.20}$ keV and 
$3.06^{+0.10}_{-0.20}$ keV.
The average estimate of the metal abundance in unit of the solar values from
Grevesse \& Sauval (1998) is $0.62 \pm 0.05$ $Z_{\odot}$ in A3528-N and
$0.45 \pm 0.04$ $Z_{\odot}$ in A3528-S.
A dramatic increase in metallicity is measured within 30$"$, 
with values of $1.79 \pm 0.20$ ($2.16^{+0.23}_{-0.19}$ after deprojection) 
$Z_{\odot}$ in the Northern clump and $1.12 \pm 0.11$
($1.41^{+0.16}_{-0.15}$) $Z_{\odot}$ in the Southern clump.
When rescaled to the  set of abundances of
Anders \& Grevesse (1989), commonly used in the past literature, 
the metallicity in the core of A3528-N is 1.21 
$Z_{\odot}$, a striking excess and together with the Centaurus cluster 
(Molendi et al. 2002; Sanders \& Fabian 2002) the largest measured so far, 
exceeding the solar value.
Their bolometric luminosities are $1.20 \times 10^{44}\,\rm{erg\,s^{-1}}$ 
for A3528-N and $1.38 \times 10^{44}\,\rm{erg\,s^{-1}}$ for A3528-S and they are
consistent within the scatter observed in L-T relation for 
nearby systems (eg Fukazawa et al. 1998; Markevitch 1998)  
\item the presence in A3528-N and A3528-S of
a centrally peaked surface brightness (see Fig.~\ref{sources}), 
a low temperature core and a steep positive gradient in 
metallicity moving inward (see Figs.~\ref{fig:prof}, \ref{map} and \ref{plotmap}), 
with a cooling time that is about 10 per cent of the age of the Universe
at $z=0.053$, suggests that they are in a relaxed state, despite
their relative small separation (0.9 $h_{70}^{-1}$ Mpc as projected 
on the sky) and large masses ($\sim 1 \times 10^{14} h_{70}^{-1} M_{\odot}$
at the virial radius) that should imply a strong merging action;
\item an extended soft X-ray emission is detected between A3528-N and A3528-S,
connecting the two clumps like a narrow bridge.
We do not observe any evidence of shock heated gas, both
in the surface brightness and in the temperature map. The surface brightness 
is not azimuthal symmetric, suggesting that A3528-N is falling from 
North--West to the South--East while A3528-S is moving from 
North--East to South--West. These facts, together with the optical light
distribution more concentrated around A3528-N, indicate that 
an off-axis (with impact parameter of about $5 r_{\rm s}$), 
post-merging (the closest cores encounter happened about 1--2 Gyrs ago) 
scenario is more plausible than a head-on pre-merger phase 
for the two subclumps.

\end{itemize}

%%--------------

\begin{acknowledgements}
The present work is based on observation obtained with \xmm, an
ESA science mission with instruments and contributions directly funded by
ESA Member states and the USA (NASA).
This work has been partially supported by the Italian Space Agency grants
ASI-I-R-105-00, ASI-I-R-037-01 and ASI-I-R-063-02, and by the Italian 
Ministery (MIUR) grant COFIN2001 ``Clusters and groups of galaxies: the 
interplay between dark and baryonic matter".
FG is grateful for the hospitality and support of ESO in Garching and wishes to
thank A. Finoguenov for the suggestion of using APM 08279+5255 observation as
background. 

\end{acknowledgements}

%%---------------

\appendix
%
%%%%%%%%%%%%%%%%%%%%%%%%%%%%%%%%%%%%%%%%%%
\section{Details on the spectral analysis}
%%%%%%%%%%%%%%%%%%%%%%%%%%%%%%%%%%%%%%%%%%

\begin{figure}[!]
\begin{center} 
\epsfig{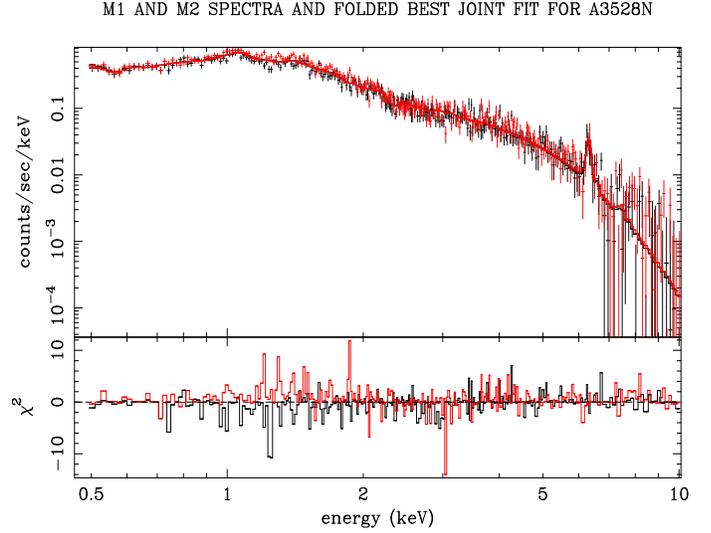}
\caption{Joint fit of the two M1 (in black) and M2 (in red) spectra of 
A3528-N. Residuals are evident in the soft (0.5-2.0 keV) band.{\label{m1m2}}} 
\end{center}
\end{figure}

\begin{figure*}[!]
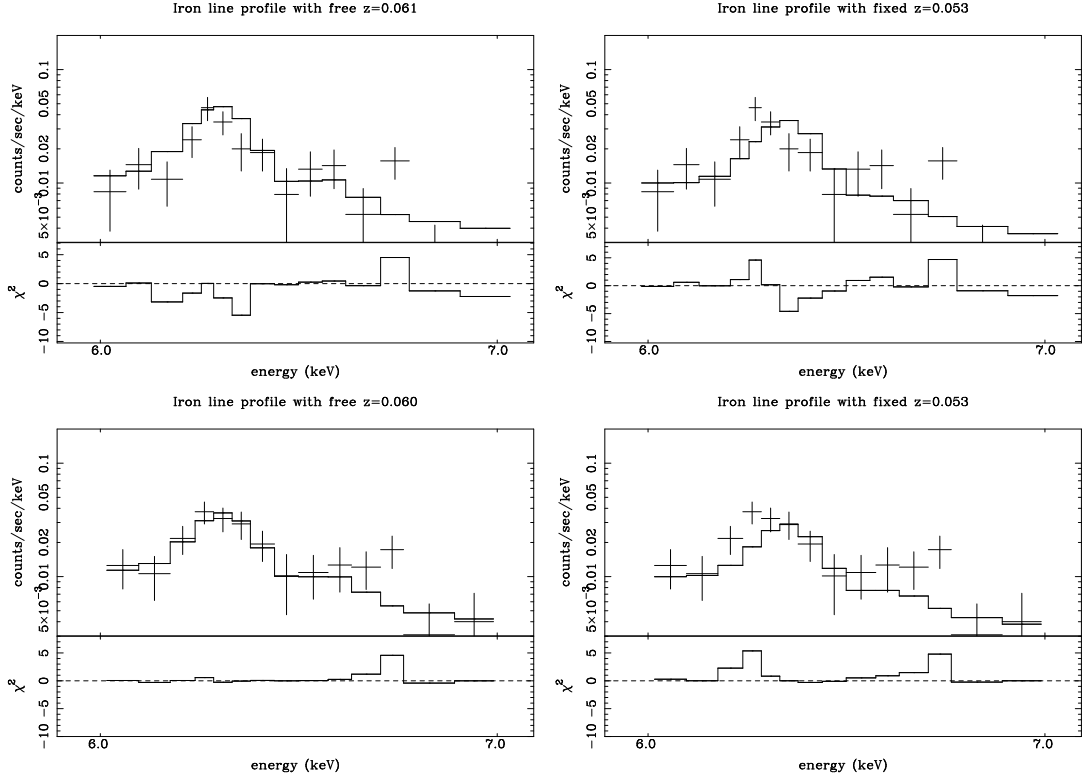
 
\begin{center}
\hspace{-0.5 truecm}{\epsfig{file=iron1.ps,width=5.0cm,height=7.0cm,angle=-90}
\hspace{0.2 truecm}\epsfig{file=iron2.ps,width=5.0cm,height=7.0cm,angle=-90}}
\vskip 0.2truecm
\hspace{-0.5 truecm}{\epsfig{file=ironfreenew.ps,width=5.0cm,height=7.0cm,angle=-90}
\hspace{0.2 truecm}\epsfig{file=ironfixednew.ps,width=5.0cm,height=7.0cm,angle=-90}}
\caption{MOS 1 data and folded model of the K iron line with free redshift 
and with fixed redshift at the optical value of 0.053 with the SASv5.3.3 in the upper panels and with SASv5.4.1 in the bottom panels.{\label{ironline}}} 
\end{center}
\end{figure*}

In the spectral fitting done with the vignetting correction factor,
we use the following response matrices:  m1\_medv9q20t5r6\_all\_15.rsp (MOS1), 
 m2\_medv9q20t5r6\_all\_15.rsp (MOS2) and, for PN,
the set of ten single-pixel matrices (from epn\_ff20\_sY0\_medium.rsp to
epn\_ff20\_sY9\_medium.rsp) depending on the mean of the values of ``RAWY''
of the photons collected in the region used to extract the spectrum.
When the arf method is applied, we use m1\_r6\_all\_15.rmf (MOS1),  
m2\_r6\_all\_15.rmf (MOS2) and as before, depending
on the mean ``RAWY'' of the region, the set of matrices 
from epn\_ff20\_sY0.rmf to epn\_ff20\_sY9.rmf for PN.

Bearing in mind that the observation was proposed to investigate
in great detail the region between the two subclusters
and the bulk of the emission of the two subclumps is
7 arcmin off-axis, where the quality of calibration is worse than
for on-axis sources, we summarize here the main problems
that emerged in the spectral analysis:

\begin{itemize}
\item there are differences between M1 and M2 at soft energies
for A3528-N, resulting in very different $\rm{N_{H}}$ and different $\rm{kT}$
from the separate fits. This can be easily see in the residuals of 
the joint fit of the two cameras in Fig.{\ref{m1m2}}.
The problem could be related to an incorrect modeling of the 
quantum efficiency of the
two detectors (Lumb, private communication): 
the same problem emerges in the our re-analysis of the same regions in 
detector coordinates of M87 observation, performed with the thin filter. 

\item problems in the CTI/gain correction for MOS cameras, 
in particular the M1 where we see a clear discrepancy in the 
redshift determination which is different from
the optical one. This could be due to the systematic drift to lower values 
(up to 20 eV at 6 keV) of the MOS line energies (Kirsch et al. 2002). 
If we fix the value of \NH to the 21cm value and the redshift to the
optical one, the M1 abundance is considerably lower ($0.54\pm0.08$)
due to the fact that it is impossible to fit correctly the iron K line 
which drives the
abundance and redshift determination, as shown in the upper panels of 
Fig.{\ref{ironline}}. This problem should be better addressed in the new 
version of the SAS, SASv5.4.1. We reprocessed the data but the problem is 
still open, although the line profile seems better determined, as shown in the 
bottom panel of Fig.{\ref{ironline}}.  

\item differences between MOS and PN cameras, due to calibration 
uncertainties or to problems of background subtraction. 
It is known that PN camera shows a up to 15\% higher flux than
MOS from 0.5-1.5 keV, while for energies above 5 keV the MOS flux is 
up to 20\% higher (Kirsch et al. 2002).
But also a not correct background subtraction could result in different 
temperature determination.  
\end{itemize}

For what concern the problems with the MOS1 camera, 
also our own analysis of the Perseus cluster, with superb statistics, 
gives systematically higher temperatures and abundances
 for MOS1 respect to MOS2 and PN.
The conclusions of a recent work aimed at assessing the EPIC spectral 
calibration using a simultaneous \xmm and \sax observation of 3C273 
strengthen this fact:
the MOS-PN cross calibration has been achieved to the available statistical 
level except for the MOS1 in the 3-10 keV band which returns flatter spectral 
slope (Molendi \& Sembay 2003).

%%%%%%%%%%%%%%%%%%%%%%%%%%%%%%%%%%%%%%%%%%%%%%%%%%%%%%%%
\section{Comparison of methods of vignetting correction}
%%%%%%%%%%%%%%%%%%%%%%%%%%%%%%%%%%%%%%%%%%%%%%%%%%%%%%%%

\begin{table*}[!]
\caption{Results of spectral fits on the two clusters for the three different cameras. 
PNS means that only single events (PATTERN==0) are considered.
In the upper half  we show the results obtained using
the effective area files, while in the bottom half the results obtained performing
the vignetting correction directly on the events.{\label{tabfit}}}
\begin{center}
\begin{tabular}{|c|c|c|c|c|c|c|}
\hline
  & $\rm{N_{H}}$ ($10^{22}\,\rm{cm^{-2}}$) & kT (keV) & Z & z & $\rm{norm}^{a}$ & $\chi^{2}$/d.o.f. \\
%\vspace{0.5cm}
\hline
A3528 N M1 & $0.0285^{+0.0059}_{-0.0062}$ & $5.04^{+0.24}_{-0.22}$ & $0.75^{+0.11}_{-0.10}$ & $0.061^{+0.003}_{-0.002}$ & $0.0115^{+0.0003}_{-0.0003}$ & 256/256 \\
A3528 N M2 & $0.0561^{+0.0060}_{-0.0057}$ & $4.16^{+0.14}_{-0.15}$ & $0.56^{+0.07}_{-0.07}$ & $0.056^{+0.002}_{-0.002}$ & $0.0138^{+0.0004}_{-0.0003}$ & 305/263 \\
A3528 N PNS & $0.0361^{+0.0033}_{-0.0032}$ & $4.22^{+0.10}_{-0.11}$ & $0.59^{+0.06}_{-0.06}$ & $0.052^{+0.002}_{-0.003}$ & $0.0118^{+0.0003}_{-0.0003}$ & 587/525 \\
A3528 S M1 & $0.0500^{+0.0052}_{-0.0051}$ & $4.52^{+0.13}_{-0.14}$ & $0.43^{+0.06}_{-0.06}$ & $0.055^{+0.002}_{-0.002}$ & $0.0147^{+0.0003}_{-0.0003}$ & 304/301 \\
A3528 S M2 & $0.0502^{+0.0052}_{-0.0051}$ & $4.25^{+0.13}_{-0.13}$ & $0.39^{+0.06}_{-0.05}$ & $0.062^{+0.002}_{-0.002}$ & $0.0148^{+0.0003}_{-0.0003}$ & 299/296 \\
A3528 S PNS & $0.0412^{+0.0030}_{-0.0037}$ & $4.40^{+0.11}_{-0.10}$ & $0.46^{+0.05}_{-0.05}$ & $0.049^{+0.005}_{-0.005}$ & $0.0139^{+0.0002}_{-0.0003}$ & 614/582 \\
\hline
\end{tabular}

\vskip0.5truecm
\begin{tabular}{|c|c|c|c|c|c|c|}
\hline
  & $\rm{N_{H}}$ ($10^{22}\,\rm{cm^{-2}}$) & kT (keV) & Z & z & $\rm{norm}^{a}$ & $\chi^{2}$/d.o.f. \\
%\vspace{0.5cm}
\hline
A3528 N M1 & $0.0317^{+0.0059}_{-0.0059}$ & $4.79^{+0.24}_{-0.17}$ & $0.70^{+0.11}_{-0.09}$ & $0.062^{+0.003}_{-0.003}$ & $0.0109^{+0.0003}_{-0.0003}$ & 341/337 \\
A3528 N M2 & $0.0581^{+0.0066}_{-0.0061}$ & $4.05^{+0.15}_{-0.17}$ & $0.57^{+0.08}_{-0.08}$ & $0.056^{+0.003}_{-0.003}$ & $0.0132^{+0.0004}_{-0.0004}$ & 404/344 \\
A3528 N PNS & $0.0307^{+0.0032}_{-0.0034}$ & $4.02^{+0.10}_{-0.11}$ & $0.58^{+0.05}_{-0.06}$ & $0.045^{+0.002}_{-0.004}$ & $0.0109^{+0.0002}_{-0.0002}$ & 720/670 \\
A3528 S M1 & $0.0485^{+0.0054}_{-0.0050}$ & $4.44^{+0.13}_{-0.15}$ & $0.39^{+0.06}_{-0.06}$ & $0.054^{+0.002}_{-0.002}$ & $0.0141^{+0.0003}_{-0.0003}$ & 367/336 \\
A3528 S M2 & $0.0530^{+0.0052}_{-0.0052}$ & $4.22^{+0.13}_{-0.13}$ & $0.39^{+0.06}_{-0.06}$ & $0.063^{+0.003}_{-0.003}$ & $0.0138^{+0.0003}_{-0.0003}$ & 338/331 \\
A3528 S PNS & $0.0414^{+0.0034}_{-0.0036}$ & $4.23^{+0.12}_{-0.11}$ & $0.44^{+0.05}_{-0.05}$ & $0.049^{+0.003}_{-0.003}$ & $0.0129^{+0.0002}_{-0.0002}$ & 770/671 \\
\hline
\end{tabular}

\end{center}
\vskip 0.4truecm
$^{a}$ The normalizations are quoted in units of $10^{-14}\,n_{e}\,n_{p}V/4 \pi D_{A}(1+z)^{2}$ as done in XSPEC. 
\end{table*} 

The reduction in effective area with radial distance from the field of view 
centre, an effect called  
vignetting, can seriously bias the brightness distribution and 
spectral modeling of extended sources as galaxy clusters.
It is now customary in EPIC analysis of X-ray clusters to perform 
the vignetting correction 
directly to spectra, weighting each photon by the inverse of its vignetting, 
which is a 
function of off-axis angle and energy (vignetting is more severe 
for high energy photons). 
This approach is similar to the one used in the analysis of ROSAT PSPC data, 
corresponding to the CORRECT mode in EXSAS (Zimmermann et al. 1998).
This method was proposed for XMM analysis of galaxy clusters by 
Arnaud et al. (2001) and opposed
to the widely used method to compare  the observed spectrum in 
a particular region of the 
detector with the incident source spectrum using an effective area (the ARF). 
This emission
weighted effective area in the regions is derived from the observed 
global photon spatial 
distribution (see Davis 2001 for a formal derivation of the ARF in
 presence of an extended source). 
This is the main drawback because, apart from introducing extra noise, 
it biases the 
determination of spectral parameters. In fact the observed photon distribution 
is always more pronounced 
towards the center than the source distribution, so too much weight is 
given to central regions
and the overall effective area is overestimated, and the overestimate 
is higher for higher
energy photons. These kinds of biases are not introduced applying the 
correction directly to 
spectra because does not introduce any a priori assumptions 
about the spatial variation 
of the source.

As a partial overcome of the problems in the ARF method
the ARF can be flux weighted
using an image divided by an exposure map, in order to better represent 
the real source
spatial distribution. This was done, following the recipe of 
Saxton \& Siddiqui (2002), using
exposure corrected images as detector maps (accumulated over square regions  
1 arcmin on each side greater than the spectra extraction region) and 
with the parameter 
\emph{extended source} of the SAS task \emph{arfgen} switched to true 
(in order to avoid the 
PSF encircled energy correction). Obviously by dividing the image by a 
unique exposure map 
the real source spatial distribution is far from being recovered, because 
of systematic 
errors introduced by applying monochromatic exposure maps 
(as it is the case of the exposure 
maps created by the SAS task \emph{eexpmap}, which assumes an energy 
which corresponds to the mean of the energy boundaries) to broad band 
images which encompasses a large range of 
variation in the effective area. These errors can be reduced if  
an exposure map is weighted according to the specific model of the incident 
spectrum (and this is easy for cluster  spectra because a thermal spectrum 
with temperature the mean temperature of the cluster can be adopted). 
If the image contains significant spectral variation 
(as in cluster images, if they are relaxed because
they have a cool core which can have a significantly lower temperature 
than the average of the cluster and if they are not relaxed because they 
can have significant temperature substructures) a single set of weights and 
thus a single exposure map cannot be applied to the entire image. Different
exposure maps and different weights should be used for different parts of the
image or the bandpass must be restricted to a range where the 
response is nearly flat (Houck 2001).

In this section as a first simple approach we compare the results obtained 
with the method
of Arnaud (2001) and the ARF method using as a source distribution the 
image in the 0.5-8.0
 keV band divided by a monochromatic exposure map with an energy of 4.25 keV (roughly
according to the mean temperature of the two clusters). 
We fit the spectra of the two clumps with the two different methods and 
making no correction
for the background and leaving all the parameters free, in order to take into 
account any 
possible difference arising. The results are shown in Tab.\ref{tabfit} and we 
can see that the spectral parameters obtained by the two methods are consistent
at the 1$\sigma$ level.

%%%%%%%%%%%%%%%%%%%%%%%%%%%%%%%%
\section{Background subtraction}
%%%%%%%%%%%%%%%%%%%%%%%%%%%%%%%%

\begin{figure*}[!] 
\begin{center}
\hspace{-0.5 truecm}{\epsfig{file=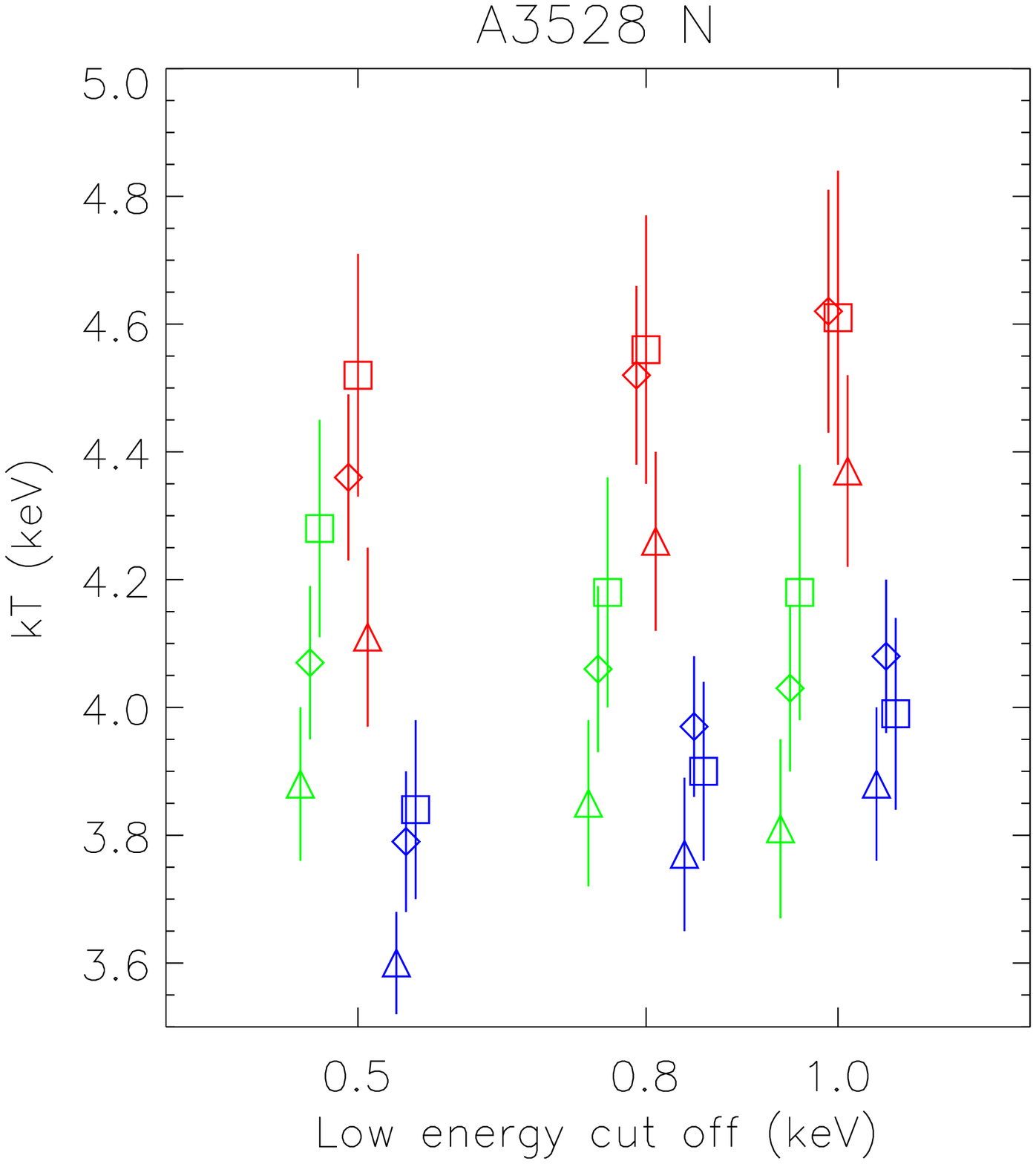,width=7.0cm,height=6.0cm}\hspace{0.2 truecm}\epsfig{file=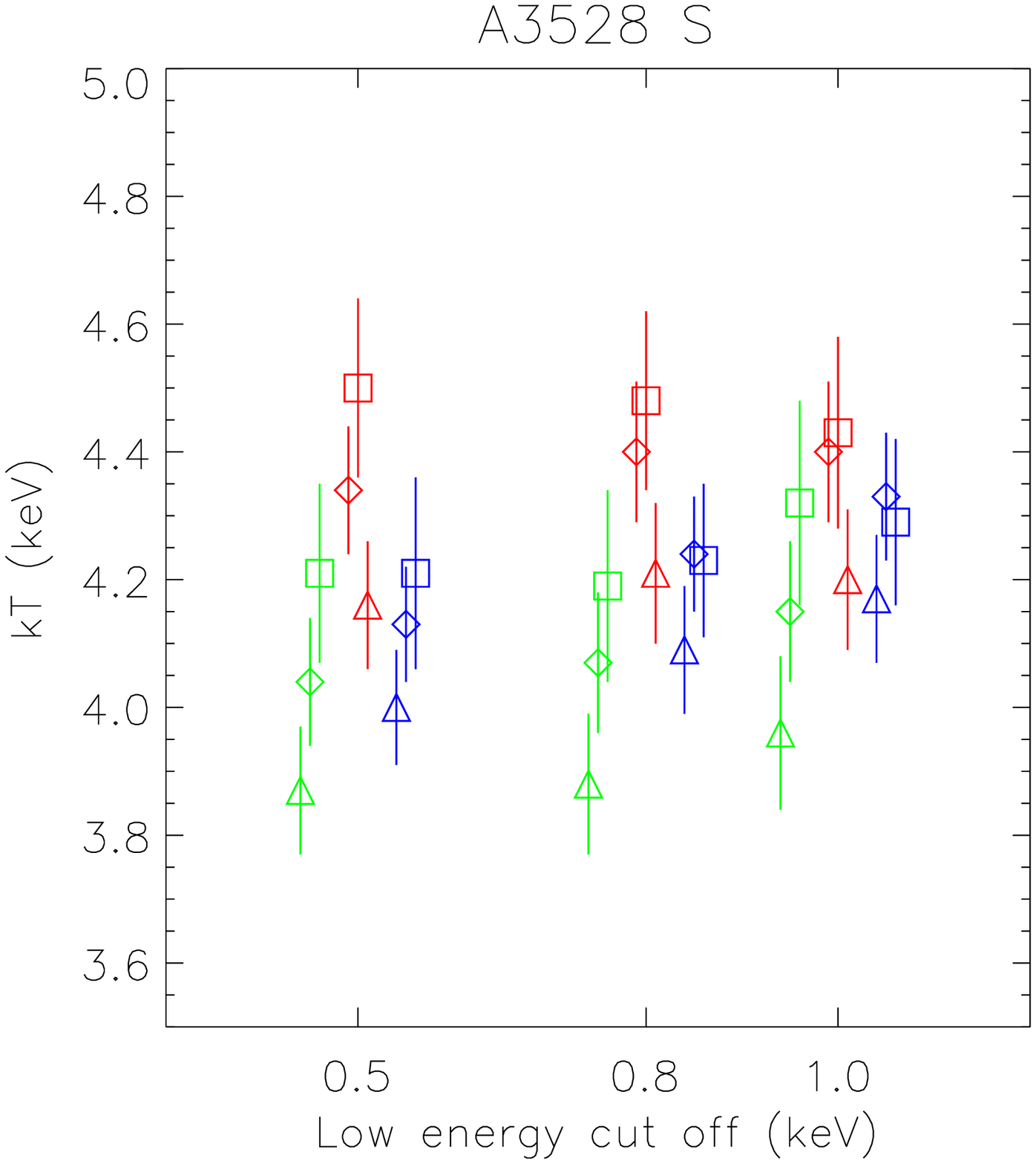,width=7.0cm,height=6.0cm}}
\caption{Measured temperature in function of different low energy cutoffs (the high energy cutoff is fixed at 8.0 keV because it had no effects on the measured temperature) for MOS1 (green), MOS2 (red) and PN singles (blue). Diamonds referred to fits with no background renormalization, triangles to fits with the renormalized background and squares to fits with the double subtraction. 
Uncertainties are at the 68\% level for one interesting parameter 
($\Delta\chi^{2}=1$).{\label{appB_bands}}} 
\end{center}
\end{figure*}

\begin{figure*}[!] 
\begin{center}
\hspace{-0.5 truecm}{\epsfig{file=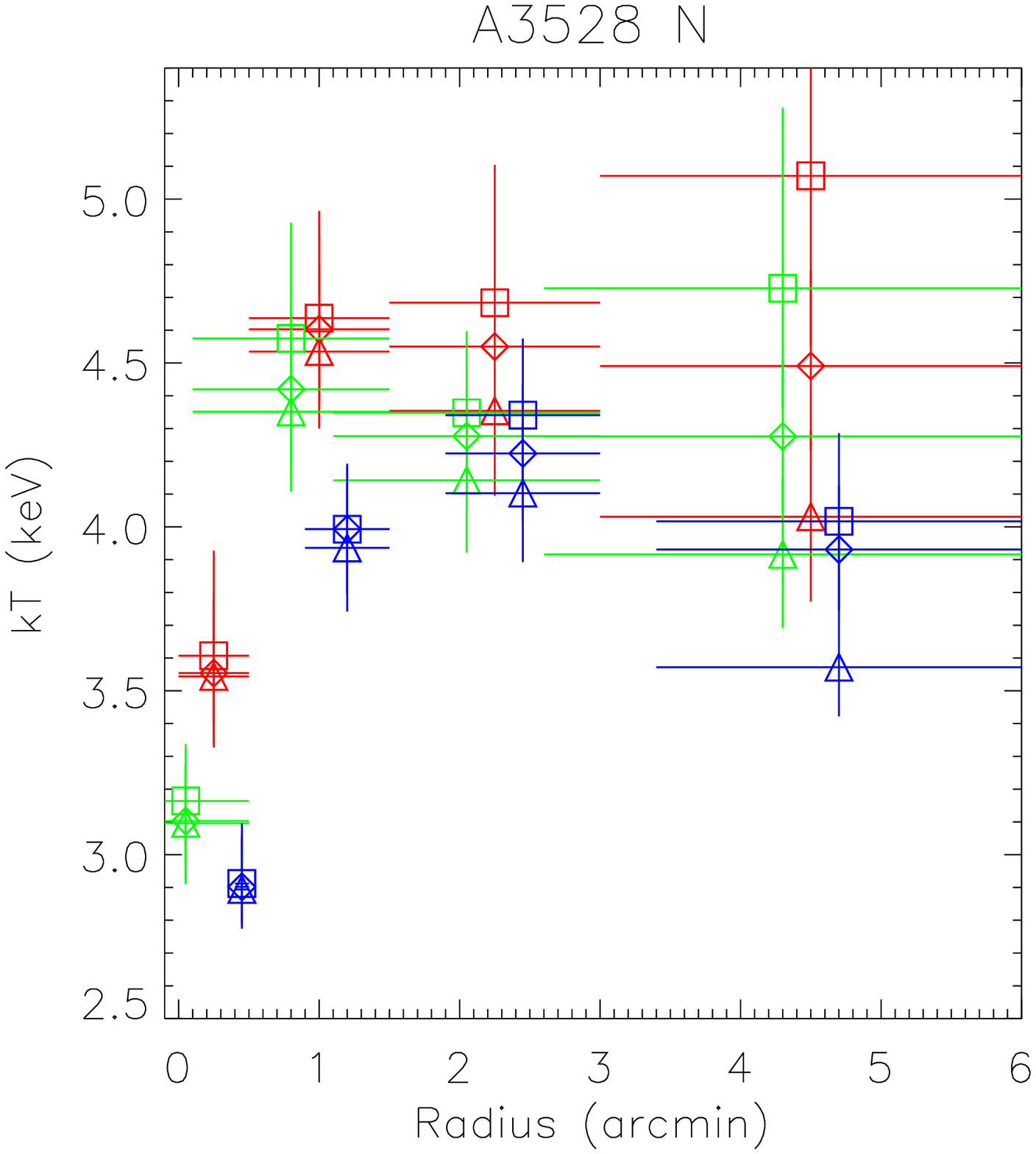,width=7.0cm,height=6.0cm}\hspace{0.2 truecm}\epsfig{file=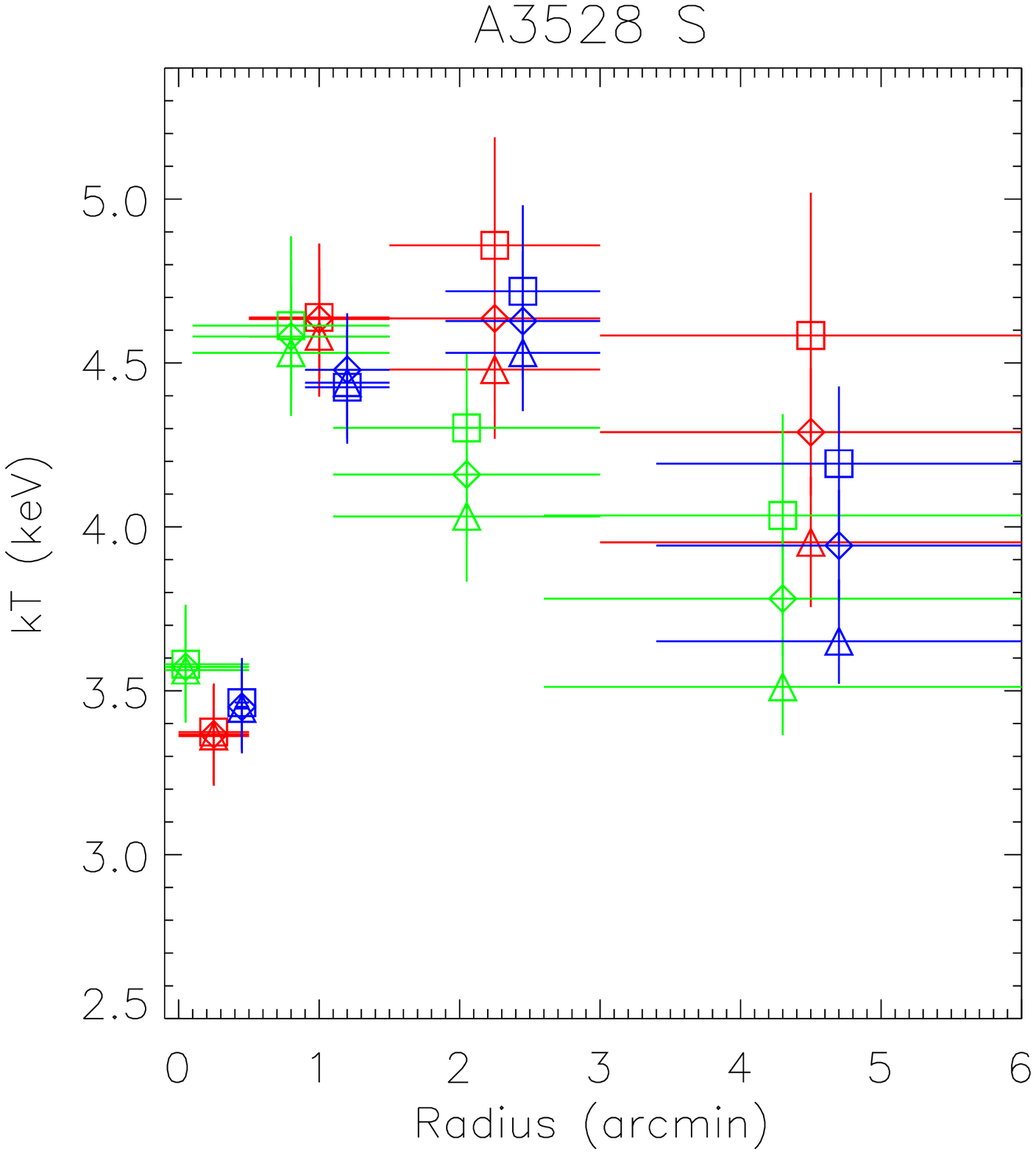,width=7.0cm,height=6.0cm}}
\caption{Same symbols as in Fig.{\ref{appB_bands}} for radial profiles.{\label{appB_radial}}} 
\end{center}
\end{figure*}

For the detail of the double subtraction method we refer the reader to 
Appendix A of Arnaud et al. (2002).
In this section we want to show how the different steps in the background 
subtraction affected the most sensitive of the spectral parameters, the
temperature.
In Fig.\ref{appB_bands} and in Fig.\ref{appB_radial} we see that the 
corrections go
in the right way, in the sense that renormalizing the background at the level
of the observation returns a lower temperature, while the further correction
for the soft background returns a higher temperature. This correction is as
we expect more crucial in the outer low surface brightness part of the cluster,
as we see in the last radial bins of Fig.\ref{appB_radial}, whether to indicate
or not a temperature decline.
A good test for the robustness of the temperature determination is to see its
variation with the choosen energy band for fitting. The results
for the subclumps are not much influenced by the fitting energy band.

\end{document}